\documentclass[numberedappendix]{emulateapj}
\usepackage{natbib}
\usepackage{epsfig}
\usepackage{amsmath}
\usepackage{color}

\slugcomment{Accepted for publication in The Astrophysical Journal}
\shorttitle {Disk-Wind Model for the NIR Excess in Protostars}
\shortauthors{Bans and K\"{o}nigl}

\begin{document}
\title{A Disk-Wind Model for the Near-Infrared Excess Emission in Protostars}

\author{Alissa Bans and Arieh K\"{o}nigl}

\affil{Department of Astronomy \& Astrophysics and The Enrico Fermi Institute, The University of Chicago, Chicago, IL 60637, USA; abans@uchicago.edu, akonigl@uchicago.edu}
\begin{abstract} 
Protostellar systems, ranging from low-luminosity classical T Tauri and Herbig Ae stars to high-luminosity Herbig Be stars, exhibit a near-infrared (NIR) excess in their spectra that is dominated by a bump in the monochromatic luminosity with a peak near $3\, \micron$. The bump can be approximated by a thermal emission component  of temperature $\sim 1500\,$K that is of the order of the sublimation temperature of interstellar dust grains. In the currently popular ``puffed-up  rim'' scenario, the bump represents stellar radiation that propagates through the optically thin inner region of the surrounding accretion disk and is absorbed and reemitted by the dust that resides just beyond the dust sublimation radius $r_{\rm sub}$. However, this model cannot account for the strongest bumps measured in these sources, and it predicts a pronounced secondary bounce in the interferometric visibility curve that is not observed. In this paper we present an alternative interpretation, which attributes the bump to reemission of stellar radiation by dust that is uplifted from the disk by a centrifugally driven wind. Winds of this type are a leading candidate for the origin of the strong outflows associated with protostars, and there is observational evidence for disk winds originating on scales $\sim r_{\rm sub}$. Using a newly constructed Monte Carlo radiative transfer code and focusing on low-luminosity sources, we show that this model can account for the NIR excess emission even in bright Herbig Ae stars such as AB Auriga and MWC~275, and that it successfully reproduces the basic features of the visibilities measured in these protostars. We argue that a robust dusty outflow in these sources could be self-limiting --- through shielding of the stellar FUV photons --- to a relatively narrow launching region between $r_{\rm sub}$ and $\sim 2\, r_{\rm sub}$. We also suggest that the NIR and scattered-light variability exhibited by a source like  MWC~275 can be attributed in this picture to the uplifting of dust clouds from the disk.
\end{abstract}


\section{Introduction}
\label{sec:intro}

Protostars ranging from low to high masses are understood to form in molecular 
cloud cores that undergo gravitational collapse. In general, the gas in the cloud possesses angular 
momentum, so a core undergoing collapse eventually encounters a centrifugal barrier and settles 
into a rotationally supported disk. It is believed that most of the mass that is assembled into young 
stars reaches the stellar surface through the associated disk \citep[e.g.,][]{CalvetEtc00}. The 
existence and properties of protostellar disks have been inferred from a variety of spectroscopic and  imaging observations \citep[e.g.,][]{MillanGabetEtc07}, whereas the arrival of disk material onto the stellar surface (where it liberates its gravitational potential energy) has been deduced from measurements of excess (over the intrinsic stellar radiation) optical/UV continuum emission \citep[e.g.,][]{HartiganEtc95,CorcoranRay98}. Protostellar spectra also exhibit a near-infrared (NIR) 
excess, which can in part be attributed to the emission from the disk surface (representing the intrinsic disk radiation as well as radiation originating in the star that is absorbed and reprocessed by the disk). However, in many cases involving both intermediate-mass Herbig Ae/Be (HAeBe) stars \citep[e.g.,][]{HillenbrandEtc92} and low-mass classical T Tauri (cTT) stars \citep[e.g.,][]{MuzerolleEtc03}, the NIR excess is dominated by a prominent ``bump'' in the monochromatic 
luminosity that peaks near $3\, \micron$ and can be approximated by a thermal emission component of temperature $\sim 1500\,$K. As reviewed in \citet[][hereafter DM10]{DMReview}, this bump cannot be reproduced by disk models that are everywhere optically thick to the stellar radiation. 

Protostellar systems generally also exhibit strong outflows that are evidently powered by the accretion process. It is commonly believed that these outflows are driven centrifugally by a large-scale, ordered magnetic field \citep[e.g.,][]{PudritzNorman83}, although it has not yet been conclusively established whether this field originates in the protostar 
\citep[e.g.,][]{ShuEtc00,RomanovaEtc09,Cranmer09}, in which case the outflow must emanate close to the protostellar surface, or in the disk \citep[e.g.,][]{KoniglPudritz00,RayEtc07}, in which case the launching region can lie further out. If the wind-driving region extends beyond the dust sublimation radius $r_{\rm sub}$ (the disk radius where the equilibrium temperature of  dust grains that are exposed to the protostellar radiation field is equal to the sublimation temperature of the grains) then disk outflows of this type offer a promising means of accounting for the observed $\sim 3\, \micron$ bump. This is because centrifugally driven winds can efficiently uplift dust  from the disk (through gas--grain collisions), which gives rise to a vertically stratified dusty gas distribution above the disk at radii $\gtrsim r_{\rm sub}$ (\citealt{Safier93a}; see Section~\ref{subsec:spectral}). The dusty outflow can in principle intercept a significant fraction of the continuum radiation emitted by the protostar and the inner disk, which is, in turn, reprocessed to infrared wavelengths. The resulting spectrum exhibits a bump near $3\, \micron$ on account of the fact that the dust sublimation temperature is $T_{\rm sub} \approx 1500\,$K, which implies a cutoff in the dust emission below $\sim 2\,\micron$, and that grains have a relatively high emissivity in the few-micron range. This was demonstrated by \citet{KK94} using the semianalytic disk-wind model of \citet{BP82} and a radiative transfer code based on the ``lambda operator'' perturbation technique. 

The \citet{KK94} work was carried out in the context of active galactic nuclei (AGNs), where a similar bump had been identified as a common NIR spectral feature \citep{EdelsonMalkan86}, and where the connection between the wavelength of the bump and the expected sublimation temperature of dust had been first noted \citep{Barvainis90}. However, in view of the basic physical principles that underlie this picture, it is clear that the ``dusty hydromagnetic disk wind'' mechanism for generating the NIR spectral bump is quite general and does not depend on the specific astrophysical context in which it is realized. Indeed, at about the same time that the AGN application was worked out,  P. Safier used a simplified disk-wind model to demonstrate that this scenario could also account for the NIR spectra of protostars. However, his results were only presented in a few conference proceedings \citep[see, e.g.,][]{Konigl96} and have not circulated through the general protostellar research community.\footnote{\citet{TambovtsevaGrinin08} also demonstrated that a dusty, centrifugally driven disk outflow can intercept and reprocess a significant fraction of the stellar radiation in this class of sources, but they did not explicitly calculate the resulting infrared spectrum.} A few years later, another model was proposed to explain the NIR spectra in protostars, namely the ``puffed-up inner rim'' scenario (\citealt{NattaEtc01}; see also \citealt{TuthillEtc01}), and this has become the most widely discussed explanation of the $\sim 3\, \micron$ bump in these sources. In this picture, the inner (gaseous) disk is by and large optically thin to the stellar radiation, so the stellar photons penetrate all the way to $r_{\rm sub}$, where they are absorbed by the dust. The absorbed radiation heats the gas at this location (the rim of the dusty disk) and causes the rim to puff up. The details of this model and of the refinements that it has undergone since its inception are presented in the DM10 review. By increasing the surface area of the disk material that reprocesses the stellar radiation, the puffed-up inner rim naturally contributes to the NIR bump emission. However,  the estimated vertical extent of the rim is evidently too small to fully account for the strongest $\sim 3\,\micron$ bumps that have been observed in either HAe \citep[e.g.,][]{TannirkulamEtc08b} or cTT \citep[e.g.,][]{AkesonEtc05} stars.

The advances made in infrared interferometry over the last decade have led to a powerful new tool for probing the origin of the NIR excess in protostars. One important early result has been the inference that the radius of an ``emission ring'' model for the interferometric visibility data scales as the square root of the stellar luminosity $L_*$ over four decades in luminosity covering cTT and HAe stars as well as low-luminosity HBe stars (\citealt{MonnierMillanGabet02}; see Figure 7 in DM10). This ``size--luminosity'' relation is consistent with the expected scaling of $r_{\rm sub}$ with $L_*$ (see Equation~(\ref{eq:r_sub})), and the normalization of the measured correlation in fact implies a temperature that is close to the expected sublimation temperatures of grains.\footnote{The scaling $r_{\rm sub}\propto L_*^{1/2}$ reflects the inverse-square dependence of the radiative flux on distance from the source and arises from the fact that the location of the sublimation radius for given dust properties is determined solely by the incident flux \citep[e.g.,][]{IvezicElitzur97}.} This result thus provides direct support for the physical picture underlying both the ``dusty hydromagnetic disk wind'' and the ``puffed-up inner rim'' scenarios. However, recent high-resolution observations of two bright HAe stars, AB Auriga and MWC~275 (=\, HD~163296), have challenged the basic rim picture by revealing the absence of a ``bounce'' in the visibility curves after their initial drop with increasing baseline \citep{TannirkulamEtc08b,BenistyEtc10}. This behavior implies that the NIR emission region in the corresponding sources  is not confined to a narrow rim but, rather, is spatially extended.

In this paper we adopt the dusty disk-wind interpretation of the NIR excess emission in protostars and develop detailed diagnostic tools that enable us to apply it to observational data. In Section~\ref{sec:formulate} we review the physics of centrifugally driven disk winds and describe our modeling setup.  We also briefly describe the Monte Carlo radiative 
transfer (MCRT) code that we constructed for this project, for which we provide full details in 
Appendix~\ref{sec:AppB}. In Section~\ref{sec:results} we present our model results for the spectral energy distribution (SED) and the visibility curve for several representative wind models and mass outflow rates. We compare the results with the data for the two aforementioned HAe stars, which have been observed with an unprecedented sub-milliarcsecond resolution \citep{TannirkulamEtc08a}. We demonstrate that, unlike the ``puffed-up inner rim'' model, the hydromagnetic disk-wind model can in principle account for both the magnitude of the $\sim 3\, \micron$ bump and the shape of the visibility curve in these sources without having to invoke any other emission component. We discuss additional aspects of this problem in Section~\ref{sec:discuss}, where we also further comment on the possible relationship between protostellar and AGN dusty disk outflows, and we then summarize in Section~\ref{sec:conclude}. The  extension of this work to higher-luminosity protostars, for which radiation-pressure effects on dust must also be taken into account, will be presented in a future publication.

\section{Formulation}
\label{sec:formulate}

\subsection{Centrifugally Driven Winds}
\label{subsec:CDW}

Hydromagnetic driving is thought to be the most likely mechanism of accelerating the 
powerful outflows observed in protostars and AGNs in view of the fact that alternative mechanisms (in particular, thermal and radiative driving) are generally too weak to account for the large momentum and energy discharges inferred in these sources. This mechanism is based on the 
presence of a large-scale, ordered magnetic field that threads the accretion disk that surrounds the central mass. In the context of protostars, this field can be naturally identified with the interstellar field that permeates the natal molecular cloud core and that is dragged in by the infalling gas once gravitational collapse sets in. Alternative possibilities are that the field is generated by a disk dynamo or, as noted in Section~\ref{sec:intro}, that it originates in the star. Magnetic stresses could launch outflows in different ways, but the most commonly invoked process is centrifugal driving, wherein disk material is flung out along magnetic field lines that are inclined at a sufficiently large angle to the disk surface --- in analogy with the motion of beads along rotating, tilted, rigid wires \citep[see][]{BP82}. As the outflowing gas climbs above the disk surface, the gradient in the magnetic pressure associated with the azimuthal field component further accelerates the flow. The disk wind is collimated by the hoop stress exerted by the azimuthal field component as well as by the magnetic tension force acting along the poloidal field component. Detailed accounts of the properties of such outflows can be found in the literature 
\citep[e.g.,][]{KoniglPudritz00,PudritzEtc07,KoniglSalmeron11}.

\subsection{Model Setup }
\label{subsec:setup}

All in all, centrifugally driven winds can be accelerated and collimated very efficiently, leading to a density distribution above the disk that is strongly stratified in the vertical direction \citep[e.g.,][]{Safier93a,Safier93b}. Such winds are generally also efficient at transporting angular momentum away from the disk surface. In fact, there is a growing number of protostellar outflows in which measured transverse velocity gradients, interpreted in terms of rotation in a centrifugally driven wind, indicate a disk wind that originates at a radius $\gtrsim 1\,$AU and transports a significant fraction of the disk angular momentum in the launching region \citep[e.g.,][]{RayEtc07,ChrysostomouEtc08,CoffeyEtc08}. The inferred values of the wind launching radius are consistent with the notion that the outflows can extend beyond the dust sublimation radius, which for bright HAe stars and their inferred dust properties (grain size $\sim 1\, \micron$; sublimation temperature $\sim 1850\,$K; see Section~\ref{sec:results}) has a magnitude $r_{\rm sub} \approx 0.2\, (L_*/40\, L_\odot)^{1/2}\,$AU (see Equation~(\ref{eq:r_sub})). There is still no direct evidence for the uplifting of dust from the inner regions of protostellar disks, but several different observations \citep[e.g.,][]{NeckelStaude95,GuethEtc03,NisiniEtc05,SmithEtc05,ChrysostomouEtc07} provide indirect support for this picture. In the case of cTT and HAe stars, there are also indications from high-resolution spectro-interferometric observations \citep[e.g.,][]{TatulliEtc07,KrausEtc08,EisnerEtc10} for Br$\gamma$-emitting outflows that originate within $r_{\rm sub}$. 

\begin{figure*}
\centering
\includegraphics[totalheight=0.2\textheight, angle=0]{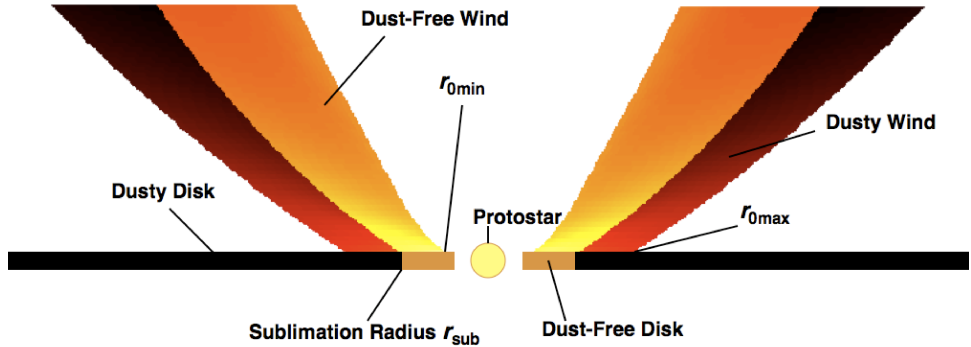}
\caption{Schematic representation of a centrifugally driven disk wind from the vicinity of a low-luminosity protostar. The gaseous wind uplifts  dust from the disk surface beyond the sublimation radius $r_{\rm sub}$ but remains dust-free closer to the protostar. The innermost radius $r_{0 \rm min}$ of the wind-launching zone may (but need not) coincide with the radius $r_{\rm m}$ (Equation~(\ref{eq:r_m})) where the disk is truncated by the stellar magnetosphere. The outer radius $r_{0 \rm max}$ of this zone is possibly limited to $\sim 2\, r_{\rm sub}$ by dust absorption of stellar FUV photons in the wind. Most of the NIR excess emission is produced by reprocessing of the stellar radiation by the dusty wind.}
\label{fig1}
\end{figure*}

In concordance with these findings, we adopt a model in which a centrifugally driven wind is launched  between an inner radius, $r_{0 \rm min}$, whose value is set to be $0.05\,$AU in all of our models, and an outer radius, $r_{0 \rm max}$,  which we take to lie at $2\, r_{\rm sub}$ (where $r_{\rm sub}$  depends on the source parameters; see Equation~(\ref{eq:r_sub})). A schematic representation of this model is shown in Figure~\ref{fig1}. The inner radius could (but need not) correspond to the disk truncation radius, which, in turn, could be determined by the stellar magnetic field. There have already been several reported measurements of a kG-strength dipolar magnetic field in HAe stars \citep[e.g.,][]{AlecianEtc09,HubrigEtc11}, although it is still unclear whether such a relatively strong field is present in only a small fraction of these sources \citep[e.g.,][]{WadeEtc2011}. Using typical values for the mass accretion rate ($\dot M_{\rm in}$) and the stellar mass ($M_*$) and radius ($R_*$), the steady-state magnetospheric truncation radius $r_{\rm m}$ corresponding to a dipolar field of equatorial surface strength $B_*$ can be estimated to be
\begin{eqnarray}
\label{eq:r_m}
r_{\rm m} \approx &0.02&\, \left (\frac{B_*}{10^3\, {\rm G}}\right )^{4/7} \left (\frac{R_*}{2.4\, R_\odot }\right )^{12/7} \left (\frac{M_*}{2.4\, M_\odot}\right )^{-1/7} \nonumber \\ &\times& \left(\frac{\dot M_{\rm in}}{10^{-7}\, M_\odot \, {\rm yr}^{-1}}\right )^{-2/7}\ {\rm AU}
\end{eqnarray}
 \citep[e.g.,][]{LongEtc05}. Our choice of $r_{0 \rm max}$ is motivated by
the argument \citep[e.g.,][]{CasseFerreira00,PesentiEtc04} that the temperature near the disk surface has to be high enough to ensure adequate mass loading of the wind. One likely source of heating for the gas in this region is stellar radiation, particularly FUV and X-ray photons, which (assuming a standard ISM gas-to-dust ratio) will reach the disk surface provided that the intervening hydrogen-nucleus column density does not exceed $\sim 10^{22}\, {\rm cm}^{-2}$ \citep[e.g.,][]{ GortiHollenbach09}. As we demonstrate in Appendix~\ref{sec:AppA}, stellar photons that reach the disk surface penetrate (for the typical parameters of our disk-wind models) through a dusty wind column $\gtrsim 10^{22}\, {\rm cm}^{-2}$ over a distance (along the disk surface) 
$\lesssim r_{\rm sub}$, which suggests that a robust outflow is unlikely to persist beyond $\sim 2\, r_{\rm sub}$. 

To carry out the radiative transfer calculation in the dusty wind, we need the gas density structure of the outflow and the dust distribution within the wind. For the density structure, we use the analytic approximations given in \citeauthor{Safier93b} (\citeyear{Safier93b}; see his Appendix~A) for several representative wind solutions that were derived semianalytically in \citet{Safier93a} using the approach of \citet{BP82}. These solutions assume a steady, axisymmetric, effectively cold, and radially self-similar disk outflow. The self-similar wind model strictly applies to an infinitely extended disk, but, given that the density structure changes mainly in the vertical direction on a scale that is much smaller than the spherical radius $R$, and that our results are not sensitive to the values of $r_{0 \rm min}$ and $r_{0 \rm max}$, we feel justified in employing them in our calculations. For the dust distribution, we simply assume a constant gas-to-dust mass ratio of 100 throughout the outflow. Although this ratio is expected to vary with location in a real disk wind, a more elaborate formulation is not warranted here in view of the fact that our spectral calculation also employs the approximation of single-size grains (see Section~\ref{subsec:spectral}) and that grains of one particular size ($\sim 1\,\micron$) indeed provide a good fit to the measured SEDs (see Section~{\ref{subsec:SED}).\footnote{One reason for why a constant gas-to-dust ratio is unlikely to apply in a real disk outflow is that, for given wind parameters, there is an upper bound on the size of grains that can be uplifted from the disk at any given radius (see Equation~(\ref{eq:a_max})). Furthermore, even grains that are dragged above the disk surface may stall and eventually fall back in if they are not small enough \citep[e.g.,][]{SalmeronIreland12}. In addition, the fact that larger grains are characterized by a higher sublimation temperature (see Equation~(\ref{eq:r_sub}) and Table~2) implies that the surface layers of the disk in the vicinity of $r_{\rm sub}$ are preferentially populated by comparatively large grains. More detailed future calculations should be able to take these effects into account.}

In the radial self-similarity formulation, any point in the wind is specified by the radius vector $\mathbf{R}$, whose cylindrical coordinates $\{r,\, \phi\, z\}$ are
\begin{equation}\label{eq:R}
\mathbf{R}=  \{r_0 \xi(\chi),\, \phi,\, r_0 \chi\}\; .
\end{equation}
Here $\chi\equiv z/r_0$ is the dimensionless vertical coordinate of a point along a streamline whose footprint intersects the disk at a radius $r_0$ (with the subscript `0' denoting the equatorial plane). All physical quantities can be represented as products of a power law in $r_0$ and a function of $\chi$. In particular, the mass density can be written as
\begin{equation}
\label{eq:rho}
\rho=\rho_{1}\left( \frac{r_0}{r_1} \right )^{-3/2} \eta(\chi)\; ,
\end{equation}
where $\rho_1$ is the density at the disk surface at the fiducial disk radius $r_1$, which we henceforth take to be $1\,$AU,
and $\eta(\chi)$ is obtained from the solution of the MHD wind equations. Using Equations~(2.1),~(2.7c), and~(5.2) in \citet{BP82}, and normalizing by values appropriate to our modeled HAe stars, we can express the fiducial density as
\begin{equation} 
\label{eq:rho1}
\begin{split}  
\rho_{1}=&3.5\times10^{-15}\left(\frac{\dot{M}_{\rm out}}{{5\times10}^{-8}\, 
M_{\odot}\, {\rm yr}^{-1}}\right)\left(\frac{M_{*}}{2\, M_{\odot}}\right)^{-1/2}  \\&\times \left( \frac{\psi_0}
{0.03} \frac{\ln{(r_{0 \rm max}/r_{0 \rm min})}}{2.5}\right )^{-1} \ {\rm g\ cm^{-3}}\; ,
\end{split}
\end{equation}
where $\dot M_{\rm out}$ is the total mass outflow rate from the disk (between $r_{0 \rm min}$ and $r_{0 \rm max}$) and $\psi_0$ is the ratio of the vertical speed $v_z$ to the Keplerian speed $v_{\rm K} = (GM_*/r_0)^{1/2}$ at the disk surface (cf. Equation~(20) in \citealt{Safier93a}). The radially self-similar wind solutions are defined by three parameters: 
$\kappa$, the normalized mass-to-magnetic flux ratio, which characterizes the mass loading of the wind; $\lambda$, the normalized total (particle and magnetic) specific angular momentum, which describes the angular momentum transport efficiency of the wind; and $\xi_0^\prime \equiv \tan{\theta_0}$ (where $\theta_0$ is the angle between the poloidal field component and the disk normal at the disk surface), which measures the initial inclination of the magnetic field lines.\footnote{In order for a cold outflow to be driven centrifugally from the surface of a Keplerian disk, $\theta_0$ must exceed $30^\circ$.} Table~1 lists the parameters of the three representative wind solutions from \citet{Safier93b} that we utilize in our calculations.
\begin{table*}
\label{tab1}
\caption{Wind Solution Parameters}
\centering
\begin{tabular}{c  c  c  c  p{1.5in} }
\hline
\hline 
Model &  $\kappa$   &  $\lambda$  &  $\xi_0^\prime$  & Notes \\ 
\hline 
C  & 0.01 & 75.43 & 1.73  & relatively fast wind, comparatively inefficient mass loading, narrow launching angle \\ 
\hline 
E  &  0.10  & 25.63 & 3.73 & relatively slow wind, efficient mass loading, wide launching angle \\
\hline
G &  0.01 & 189.34 & 3.73 & fastest wind model, comparatively inefficient mass 
loading, wide launching angle \\
\hline
\end{tabular}
\end{table*}

If the wind transports away all the excess angular momentum of the accreted gas, and the mass accretion rate through the disk, $\dot M_{\rm in}$, remains nearly constant with radius, then the ratio of $\dot M_{\rm out}$ to $\dot M_{\rm in}$ can be approximated by $\ln\{{(r_{0\rm max}/r_{0\rm min})}/[2(\lambda -1)]\}$ \citep[see][]{KoniglSalmeron11}.  According to this estimate, $\dot M_{\rm out}/\dot M_{\rm in}$ should lie in the range $\sim 1-5\,\%$ for the models listed in Table~1. This is consistent with the observational determinations of this ratio in low-mass protostars. However, using our best-fit values of $\dot M_{\rm out}$ for the two HAe stars that we model in this paper (which are $\gtrsim 4\times 10^{-8}\, M_\odot\, {\rm yr}^{-1}$; see Figure~\ref{fig3}), the above expression would imply mass accretion rates that are in excess of those that have been inferred observationally for these two sources (which are $\lesssim 10^{-7}\, M_\odot\, {\rm yr}^{-1}$; see below). The discrepancy could be reduced if we have overestimated $\dot M_{\rm out}$ (for example, because of our neglect of other likely contributions to the NIR excess emission, as discussed in Section~\ref{sec:results}, or because we have overestimated the ratio $r_{0\rm max}/r_{0\rm min}$ in Equation(\ref{eq:rho1})). Note in this connection that a mass outflow rate $\sim 1\times 10^{-8}\, M_\odot\, {\rm yr}^{-1}$ was inferred in MWC~275 from observations of Herbig-Haro knots in the bipolar outflow emanating from this source \citep{WassellEtc06}. It is also conceivable that the mass accretion rate in these intermediate-mass sources undergoes a measurable decreases with radius because of a considerable mass loss to the wind, so that $\dot M_{\rm in}$, which is determined from observations near the stellar surface, could be significantly smaller than the local mass accretion rate near $r_{\rm sub}$. (This situation is thought to possibly occur also in AGNs; e.g., \citealt{ElitzurShlosman06}.) Intrinsic variability in the source could further contribute to the apparent discrepancy. For example, MWC~275 exhibited a NIR flux increase that did not simply correspond to the behavior of accretion diagnostics such as the Ca II triplet lines \citep{SitkoEtc08}. If the NIR variability is associated with changes in the outflow rate near $r_{\rm sub}$, which could, in turn, be related to localized variations in the accretion rate at that  radius (see Section~\ref{sec:discuss}), then it is entirely plausible that the inferred values of $\dot M_{\rm out}$ and of $\dot M_{\rm in}$ (which, as we noted above, is determined at a much smaller radius) would not obey the above scaling relation  even if they did in a strictly steady state.

Apart from the preceding considerations, the determination of the mass accretion rates in HAe and HBe stars is complicated by the fact that these objects are intrinsically bright at the same UV wavelengths where the emission from an accretion shock on the stellar surface is expected to peak. Furthermore, in contrast with cTT stars, it is not clear that the commonly invoked magnetospheric accretion model is generally applicable to the higher-mass protostars, and there is also a relative dearth of metal absorption lines available for measurements of veiling (an accretion diagnostic). In addition, Br$\gamma$ emission, which is often used to infer $\dot M_{\rm in}$ in lower-luminosity protostars, can have a significant contribution from an outflow component in many HAeBe stars \citep[e.g.,][]{TatulliEtc07,KrausEtc08,WeigeltEtc11}. The prevailing uncertainties are reflected in the range of values of $\dot M_{\rm in}$ that have been estimated for the two HAe stars under consideration. For example, in the case of MWC 275, \citet{GarciaLopezEtc06} inferred a value of $7.6\times 10^{-8}\,M_\odot\, {\rm yr}^{-1}$ based on a measurement of the Br$\gamma$ line luminosity. However, even though \citet{EisnerEtc10} measured a similar Br$\gamma$ luminosity, they deduced a higher value of $\dot M_{\rm in}$ ($\simeq 1.75\times 10^{-7}\, M_\odot\, {\rm yr}^{-1}$) on account of their use of a different interpretive scheme. \citet{DonehewBrittain11}, in turn, inferred $\dot M_{\rm in} \approx 6.9\times10^{-8}\, M_\odot\, {\rm yr}^{-1}$ for this source by using the so-called Balmer discontinuity as a veiling tracer, but \citet{MendigutiaEtc11}, employing a similar approach, deduced only an upper limit of $\sim 3.1\times 10^{-8}\, M_\odot\, {\rm yr}^{-1}$. In the case of AB Aur, \citet{GarciaLopezEtc06} estimated $\dot M_{\rm in}\approx 1.4\times 10^{-7}\, M_\odot\, {\rm yr}^{-1}$, whereas \citet{DonehewBrittain11} inferred a value of only $\sim 1.8\times 10^{-8}\,M_\odot\, {\rm yr}^{-1}$.\footnote{Note in this connection that, by attributing the origin of the hydrogen infrared lines in AB Aur to a wind, \citet{NisiniEtc95} estimated a mass \emph{outflow} rate of $\sim 3.3\times 10^{-8}\, M_\odot\, {\rm yr}^{-1}$ in this source.}

\subsection{Spectral Calculations}
\label{subsec:spectral}

All the spectral results presented in this paper were obtained with the help of a newly developed and fully tested MCRT code that is based on the radiative equilibrium and temperature correction scheme discussed in \cite{BjorkmanWood01}; it is described in detail in Appendix~\ref{sec:AppB}. Although this code is technically  2D, it fully captures the actual 3D trajectories of the modeled photons under the postulated axisymmetry. In any given run of the code, the grains are assumed to have one particular size: either small ($0.001\,\micron$), intermediate ($0.1\, \micron$), or large ($1.0\, \micron$). As discussed in \citet{Safier93a}, there is a maximum size $a_{\rm max}$ of grains at any given radial location for which the wind has sufficient momentum to overcome tidal gravity and uplift the dust from the disk surface. Combining Equation~(E7) in that paper with Equation~(\ref{eq:rho1}) above, we estimate
\begin{equation} 
\label{eq:a_max}
\begin{split} 
a_{\rm max}(r_0)&\approx 0.3 \left(\frac{\dot{M}_{\rm out}}{{5\times10}^{-8}\, 
M_{\odot}\, {\rm yr}^{-1}}\right)\left(\frac{M_{*}}{2\, M_{\odot}}\right)^{-1/2} \left (\frac{\psi_0}
{0.03}\right ) \\&\times \left (\frac{\rho_{\rm s}}{3\, {\rm g\, cm}^{-3}}   \frac{\ln{\frac{r_{0 \rm max}}{r_{0 \rm min}}}}{2.5}\right )^{-1} \left (\frac{r_0}{0.3\, {\rm AU}}\right )^{-1/2} \ \micron\; ,
\end{split}
\end{equation}
which indicates that grains much larger than $\sim 1\, \micron$ are unlikely to be present in typical disk outflows from HAe sources. (Although our model neglects dynamical interactions of dust particles in the wind, such interactions are not expected to lead to measurable growth after the grains enter the outflow.) 

We assume that the dust consists of pure silicates and we employ grain optical properties (derived under the assumption that the grains are compact and spherical) from the tabulation of  \citet{WeingartnerDraine01}.\footnote{Available at www.astro.princeton.edu/$ \sim$draine/dust/ dust.diel.html.} These properties depend on the grain size. Under conditions of radiative equilibrium, a given grain emits as much energy per unit time as it absorbs. The dust optical properties at the wavelengths where the stellar emission peaks determine how effectively this radiation is absorbed, whereas the corresponding properties at the longer wavelengths where most of the dust emission occurs determine how effectively the absorbed radiation is reemitted. Grains of size $a \ll 1\, \micron$ typically have a low cooling efficiency, whereas grains whose size exceeds a few microns have an effectively gray opacity and reradiate the absorbed stellar radiation with a comparatively high efficiency (see DM10). Since the radiative equilibrium condition determines the dust temperature, large grains tend to be cooler than small grains for a given incident flux. By the same token, large grains attain a given equilibrium temperature at a smaller radius (where the incident stellar flux is higher) than small grains. This is quantified by the scaling function $H$ in the expression
\begin{equation} 
\label{eq:r_sub}
r_{\rm sub}=  H(T_*,T_{\rm sub},a) (L_{*}/L_{\odot})^{1/2} T_{\rm sub}^{-2} \ {\rm AU}
\end{equation}
for the sublimation radius, which is obtained from the radiative equilibrium condition by substituting $T_{\rm sub}$ for the dust temperature. Table~2 lists the values of $H(T_*,T_{\rm sub},a)$ for our representative grain sizes and for three values of the dust sublimation temperature, assuming that the stellar photosphere is characterized by an effective temperature $T_*\approx 9650\,$K.

 The dust sublimation temperature depends on the grain composition as well as on the ambient gas pressure. Recent applications in the literature have often used the empirical relation presented by \citet{IsellaNatta05}, which gives $T_{\rm sub}$ for silicates as a function of the local gas density. This relation is based on the results of \citet{PollackEtc94}, who fitted a Clausius-Clapeyron equation to data for Olivine. The extraction of a density dependence from this equation was done under the implicit assumption that the dust and gas temperatures are equal, but this is typically not the case for the comparatively 
low-density disk winds that we consider. In view of this, and in the interest of simplicity, we neglect any density dependence of the sublimation temperature in our treatment and content ourselves with using a single sublimation temperature (having one of three representative values) in any given run.\footnote{We have, however, verified that our best-fit models do not significantly change if we employ instead a density-dependent sublimation temperature using the expression derived by \citet{IsellaNatta05}.} Given that previous inferences of the sublimation temperature, made in the context of the inner rim model, have ranged from $\sim 1150\,$K  for MWC~275 \citep{BenistyEtc10} to $\sim 1950\,$K for AB Aur \citep{TannirkulamEtc08b}, we adopt the following representative values for $T_{\rm sub}$: $1250\,$K, $1500\,$K, and $1850\,$K. As Table~2 demonstrates, the parameter dependence of $r_{\rm sub}$ is not, in general, monotonic: For the given value of $T_*$ and with $T_{\rm sub}$ fixed, the value of $r_{\rm sub}$ peaks for the intermediate-size grains.

In our formulation, we represent the star in a simplified fashion as a point source at the origin of the coordinate system. We further simplify the treatment by taking the disk to be a flat surface in the equatorial plane. In this limit, no stellar photons impinge on the disk directly. Radiation does, however, reach the disk as a result of scattering and reemission of stellar photons in the dusty wind. We treat the interaction of this radiation with the disk in a manner analogous to the two-zone approximation described in \citet{ChiangGoldreich97}. In this picture, the incident radiation is fully absorbed in a dusty surface layer whose temperature is calculated in the same way as that of the grains in our wind model. The heated surface layer reemits the incident radiation at longer wavelengths, and it is assumed that half of this radiation is directed outward and the other half is absorbed and thermalized in the disk interior, which then reemits it as a blackbody. We neglect, however, any potential contribution to the disk emission from energy dissipated in the accretion process. Our MCRT code follows the paths of photons that are emitted by the disk in the same manner that it handles the transfer of the stellar radiation through the wind.

In this paper we concentrate on the dusty wind and disk region beyond the sublimation radius, and our spectral calculations initially neglect any contribution of the disk or the wind at smaller radii. We nevertheless return to address this issue in Section~\ref{subsec:interior}, where we estimate in an approximate manner the effects of the thermal emission from the interior wind (due mainly to intrinsic ambipolar diffusion heating) and disk (due to reradiation of directly impinging stellar photons). We do not, however, include the effect of ``back warming''  of the interior disk and wind region (or of the star) by photons that originate beyond $r_{\rm sub}$.

\begin{figure*}
\centering
 \includegraphics[totalheight=0.5\textheight, angle=270]{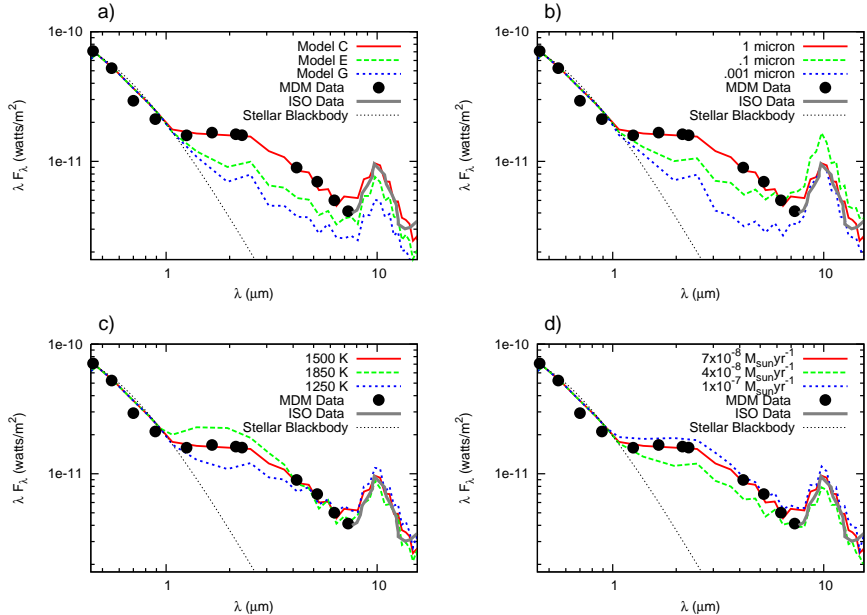}
\caption{SED fits for AB Auriga, showing the effects of varying different model parameters. In each of the four panels, the solid (red) curve represents the base model, whereas the dashed (green) and dotted (blue)  curves show the changes to the predicted spectrum that are induced by varying the value of one of the parameters: the wind model (a), the grain size (b), the sublimation temperature (c), and the mass outflow rate (d). The data points are from the MDM telescope (filled circles; \citealt{TannirkulamEtc08b}) and from {\it ISO} [thick solid (gray) line; \citealt{MeeusEtc01} and references therein]. The stellar emission  is represented by a blackbody spectrum (thin dotted line).}
\label{fig2}
\end{figure*}

\
\section{Results}
\label{sec:results}

In this Section we present the results of our spectral calculations,  employing different disk-wind models, mass outflow rates, grain sizes, and dust sublimation temperatures to infer the dependence of the NIR observational signatures on the relevant physical parameters  and to deduce the parameter values that best fit the data for AB Aur and MWC 275. We  first consider the dusty wind and disk region at $r\ge r_{\rm sub}$ and describe, in turn, the predicted spectral energy distributions (SEDs) and interferometric visibility curves , and we then discuss the expected contribution from the gas in the disk and the outflow interior to $r_{\rm sub}$. 

\begin{figure*}
\label{fig3}
\begin{center}
\includegraphics[width=0.34\textwidth, angle=270]{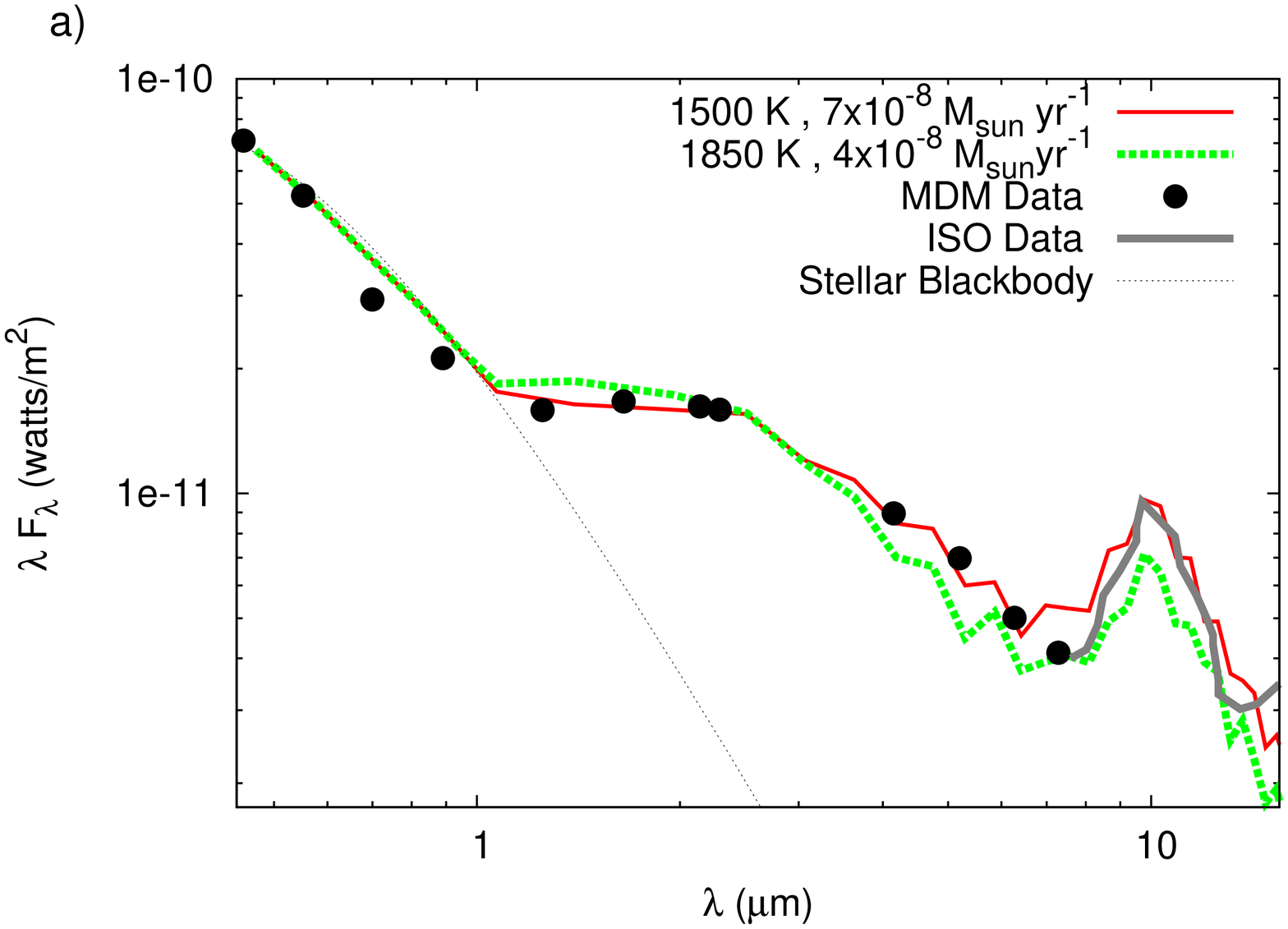}
\includegraphics[width=0.34\textwidth, angle=270]{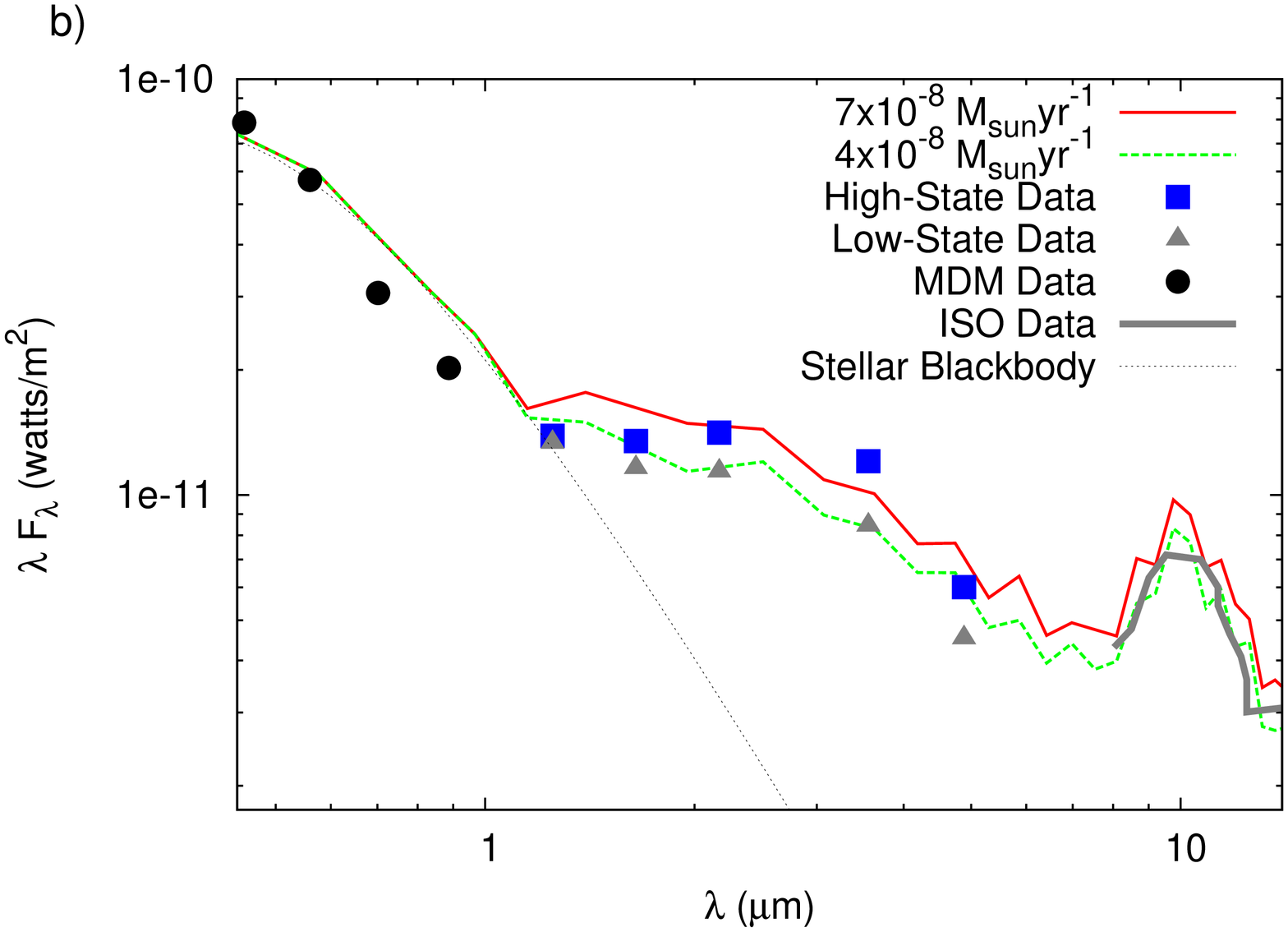}
\end{center}
\caption{Best-fit curves to the SEDs of AB Auriga (a) and MWC~275 (b). In the case of AB Aur, the two curves demonstrate the possible parameter degeneracy that can be encountered by fitting the SED data alone. In the case of MWC~275, the two curves (which both correspond to $T_{\rm sub} = 1500\,$K) demonstrate how a change in the wind outflow rate could potentially account for the NIR variability exhibited by this source. The high-state (squares)  and low-state (triangles) measurements  for MWC 275 are from \citet{SitkoEtc08}. The references for the data points from the MDM telescope and from {\it ISO} are given in the caption to Figure~\ref{fig2}.}
 \end{figure*}

\subsection{Model SEDs}
\label{subsec:SED}

\begin{table}[t]
\label{tab2}
\caption{Scaling function $H$ in Equation~(\ref{eq:r_sub}) }
\centering
\begin{tabular}{c  c c  c  p{.75 in} }
\hline
\multicolumn{4}{c }{ $T_{\rm sub}$(K)   \ \ \ \ \ ~~~~~~  $a$ ($\micron$)} \\
   & & 0.001   &  0.1   & ~~1.0 \\
\hline
&1250~~~ & 2.18e5 & 2.95e5& 1.23e5  \\
&1500~~~ & 2.15e5& 2.85e5 & 1.15e5 \\
&1850~~~ &  2.05e5& 2.65e5 & 1.09e5\\
\hline
~
 \end{tabular}
\end{table}

As was already found in previous studies \citep[e.g.,][]{IsellaEtc06,TannirkulamEtc08b}, the dust properties that most directly affect the appearance of the NIR bump are the size of the grains ($a$) and the sublimation temperature ($T_{\rm sub}$). Besides these two parameters, we consider also the dependence of the calculated SEDs on the specific disk-wind model (C, E, or G; see Table~1) and on the mass outflow rate ($\dot M_{\rm out}$). To fully specify a solution, we also need the values of the stellar mass ($M_*$) , total luminosity ($L_*$), and effective temperature ($T_*$), which are inferred observationally for any given source,\footnote{We adopt the values given in \citet{GarciaLopezEtc06}: $M_* =2.4\, M_\odot$,  $T_*=9840\,$K, $L_*=48\,L_\odot$ for AB~Aur, and $M_* =2.3\, M_\odot$,  $T_*=9450\,$K, $L_*=36\,L_\odot$ for MWC~275.} as well as the values of the parameter $\psi_0$ and of the ratio $r_{0 \rm max}/r_{0 \rm min}$ (see Equation~(\ref{eq:rho1})) --- we fix the value of $\psi_0$ to be 0.03 and evaluate the ratio of the outermost and innermost wind radii using the prescription outlined in 
Section~\ref{subsec:setup}.

Figures~\ref{fig2} and~\ref{fig3} show results from our model calculations as well as data points for AB~Aur and MWC~275.  Since our interest centers on the NIR spectral regime, the SEDs are plotted only up to wavelengths of $\sim 10\, \micron$. It is immediately apparent that this model can account for the prominent NIR bumps exhibited by these sources while also roughly reproducing the overall shapes of the measured spectra. The main contribution to the NIR excess emission comes from the dusty wind, which intercepts the stellar radiation in a ``reprocessing surface'' whose effective extent (for a given wind model) depends mainly on the mass outflow rate.  While this surface is analogous to the puffed-up inner rim, it can be significantly larger for a sufficiently high value of $\dot M_{\rm out}$; the wind model can therefore explain even the strong $\sim 3\,\micron$ bumps seen in these bright HAe sources, for which the rim model evidently falls short. Our spectral model also includes a dusty disk component, which begins to make a significant contribution to the spectrum at wavelengths $\gtrsim 4\, \micron$. Since our disk model is rather simplified (especially in its assumption of geometrical flatness), the predictions at those longer wavelength are not expected to be accurate. As the spectral regime beyond the NIR bump is not the focus of this work, we have not attempted to improve on our disk emission model, but we note that more realistic models of passive flared disks can reproduce the observed spectra at these wavelengths \citep[e.g.,][]{ChiangEtc01}.   

The dependence of the synthetic spectra on the different model parameters is illustrated in Figure~\ref{fig2} with fits to the data from AB~Aur. The base model corresponds to $1\,\micron$ grains with a sublimation temperature of $1500\,$K that are embedded in a wind that has a mass outflow rate of $7\times 10^{-8}\, M_\odot \, {\rm yr}^{-1}$ and is described by wind model~C, and each of the four panels presents variations in one of these attributes. Panel (a) shows the dependence of the spectra on the choice of wind model. The dominant effect on the emitted flux level appears to come from the   self-similarity parameter $\xi_0^\prime$, which characterizes the initial opening angle of the poloidal streamlines (see Section~\ref{subsec:setup}): The NIR flux is measurably higher in model C, which has the narrowest launching angle, but there is not much difference in the spectral properties of models E and G, which have disparate values of the self-similarity parameters 
$\kappa$ and $\lambda$ but the same (higher) value of $\xi_0^\prime$ (see Table~1). Panel (b) examines the dependence on the grain size $a$. As was noted in Section~\ref{subsec:spectral}, large ($a \gtrsim 1\, \micron$) grains cool with a comparatively high efficiency, which explains why the wind contribution to the $\sim 3 \,\micron$ bump increases with $a$. The large grains, with their effectively gray opacity, also better match the flat shape of the spectrum in the bump region. It was further noted in Section~\ref{subsec:spectral} that the value of the sublimation radius is largest for the intermediate-size ($a=0.1\, \micron$) grains (see Equation~(\ref{eq:r_sub}) and Table~2).  This implies that, for these grains, the emission at any given wavelength originates further out in the disk than it does for the other grain sizes. At those larger distances, the wind density is lower because of the $r^{-3/2}$ scaling of $\rho$ (see Equation~(\ref{eq:rho})) and of the inverse linear dependence of the normalization $\rho_1$ on $\ln({r_{0\rm max}}/r_{0\rm min})$ (see Equation~(\ref{eq:rho1})). Therefore, the fraction of the stellar radiation that is intercepted by the wind and scattered down into the disk is lower in this case, resulting in a cooler disk that contributes more to the flux at longer wavelengths. This explains why the intermediate-size grains exhibit the strongest 
$\sim 10\,\micron$ emission feature. 

Panel (c) of Figure~\ref{fig2} depicts how the predicted spectra are modified when the value of $T_{\rm sub}$ is changed. Since a higher sublimation temperature implies a smaller value of the sublimation radius, the dusty emission region moves inward to where the wind density is higher (see equations~(\ref{eq:rho}) and~(\ref{eq:rho1})), leading to a higher optical depth in absorption and hence to more reprocessing of the stellar radiation. Therefore, we can expect stronger NIR emission for higher values of $T_{\rm sub}$. This expectation is borne out in the figure, which shows that the flux becomes progressively higher as $T_{\rm sub}$ is increased. A similar trend is expected when the mass outflow rate goes up, in view of the linear dependence of the density on $\dot M_{\rm out}$ (see Equation~(\ref{eq:rho1})). This expectation, too, is confirmed by the explicit calculation, as is shown in panel (d).

Figure~\ref{fig3} presents our ``best-fit'' SED models for AB Aur and MWC~275. Because of the limited number of runs of the MCRT code that we have performed, our chosen parameter values are only accurate, on average, to $\sim 10\, \%$. Furthermore, as has already been known from previous work in this area (see DM10), a fit to the SED does not uniquely fix the underlying physical model. We illustrate this fact within the framework of our scheme in panel (a) of this figure, where we show two comparable fits to the spectrum of AB Aur that are based on different combinations of the parameters $\dot M_{\rm out}$ and  $T_{\rm sub}$. As we demonstrate in Section~\ref{subsec:visibility}, this parameter degeneracy can be lifted by considering also the best fit to the visibility data.

Panel (b) of Figure~\ref{fig3} shows SED fits for MWC~275. This source exhibited a
$\sim 30\,\%$ variability in the $1-5\, \micron$ flux level around 2002 \citep{SitkoEtc08}, and the data points associated with this ``outburst'' (the ``high'' state) and with the quiescent emission (the ``low'' state) are indicated in the figure. We fit the ``high'' state of this source with the base model of Figure~\ref{fig2}, whereas an approximate match to the `low'' state can be obtained by reducing the mass outflow rate in this model by $\Delta \dot{M}_{\rm out} \approx 3 \times 10^{-8}\, M_\odot\, {\rm yr}^{-1}$. 
Physically, the outburst exhibited by this source could be associated with the injection of a density inhomogeneity (a ``cloud'' or ``clouds'') into the wind. Such clouds would move across the line of sight at approximately the wind speed and could in principle also account, through shadowing, for the variability that was measured in this source in the scattered-light image of the outer disk (whose spatial scale is a few hundred AU) on a time scale of a few years \citep{WisniewskiEtc08}. Further discussion of this scenario is given in Section~\ref{sec:discuss}.

Our model fits in this subsection are based exclusively on the contributions of the wind and disk beyond $r_{\rm sub}$. If the inner disk is optically thin to the stellar radiation then an additional contribution from the heated inner edge of the dusty disk, as envisioned in the ``puffed-up inner rim'' scenario, could be expected. The contribution of the latter component to the NIR excess in these sources would be subordinate to that of the dusty wind, but it could conceivably help improve the detailed fit to the measured spectra. A further contribution to the observed spectrum could come from the disk and wind at $r<r_{\rm sub}$, as discussed in Section~\ref{subsec:interior}. Any additional contribution of this type to the NIR excess emission would reduce the inferred value of $\dot M_{\rm out}$ for the wind component.

\begin{figure*}  
\begin{center}    
\includegraphics[width=0.34\textwidth,angle=270]{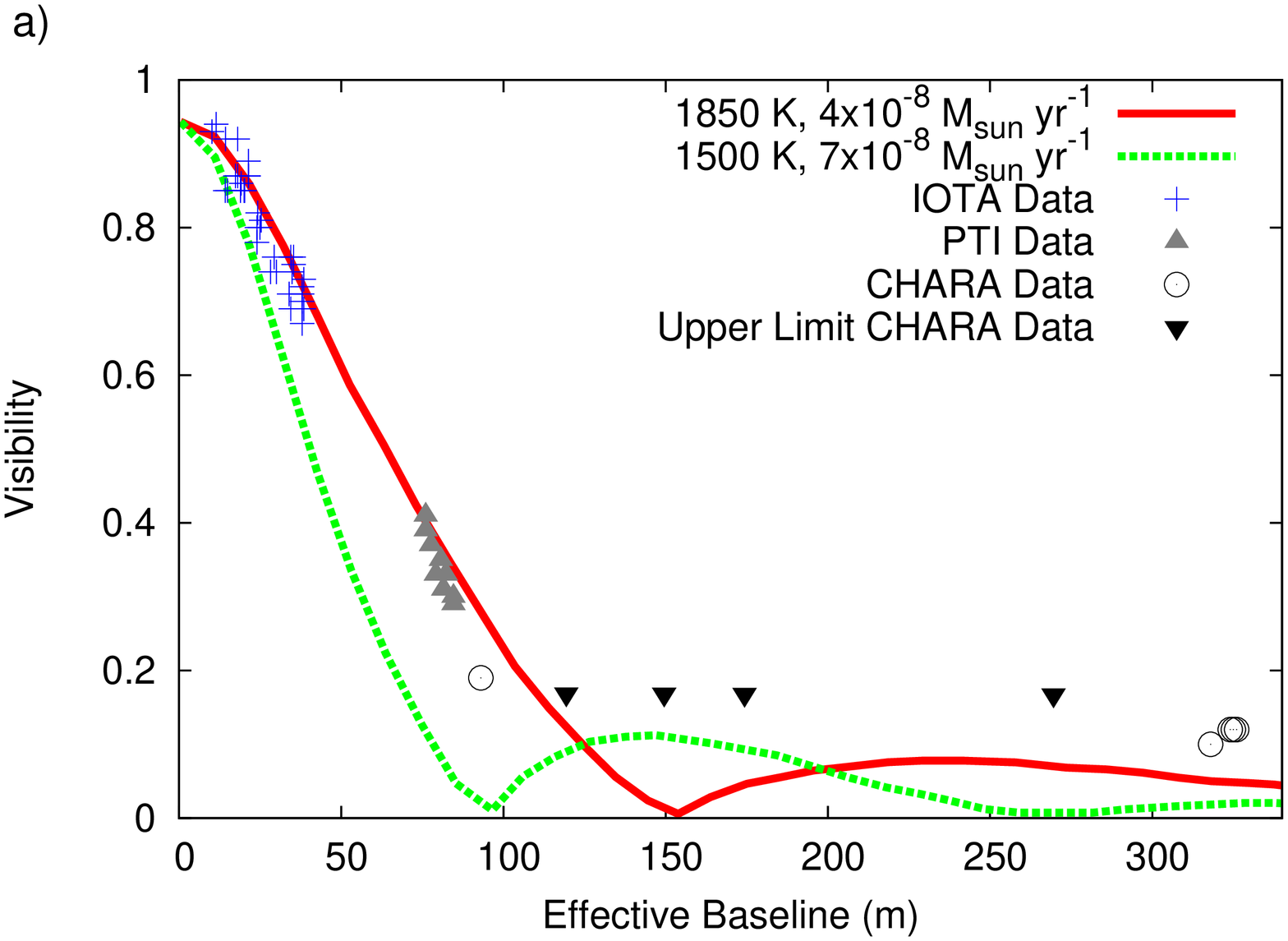} 
\includegraphics[width=0.34\textwidth,angle=270]{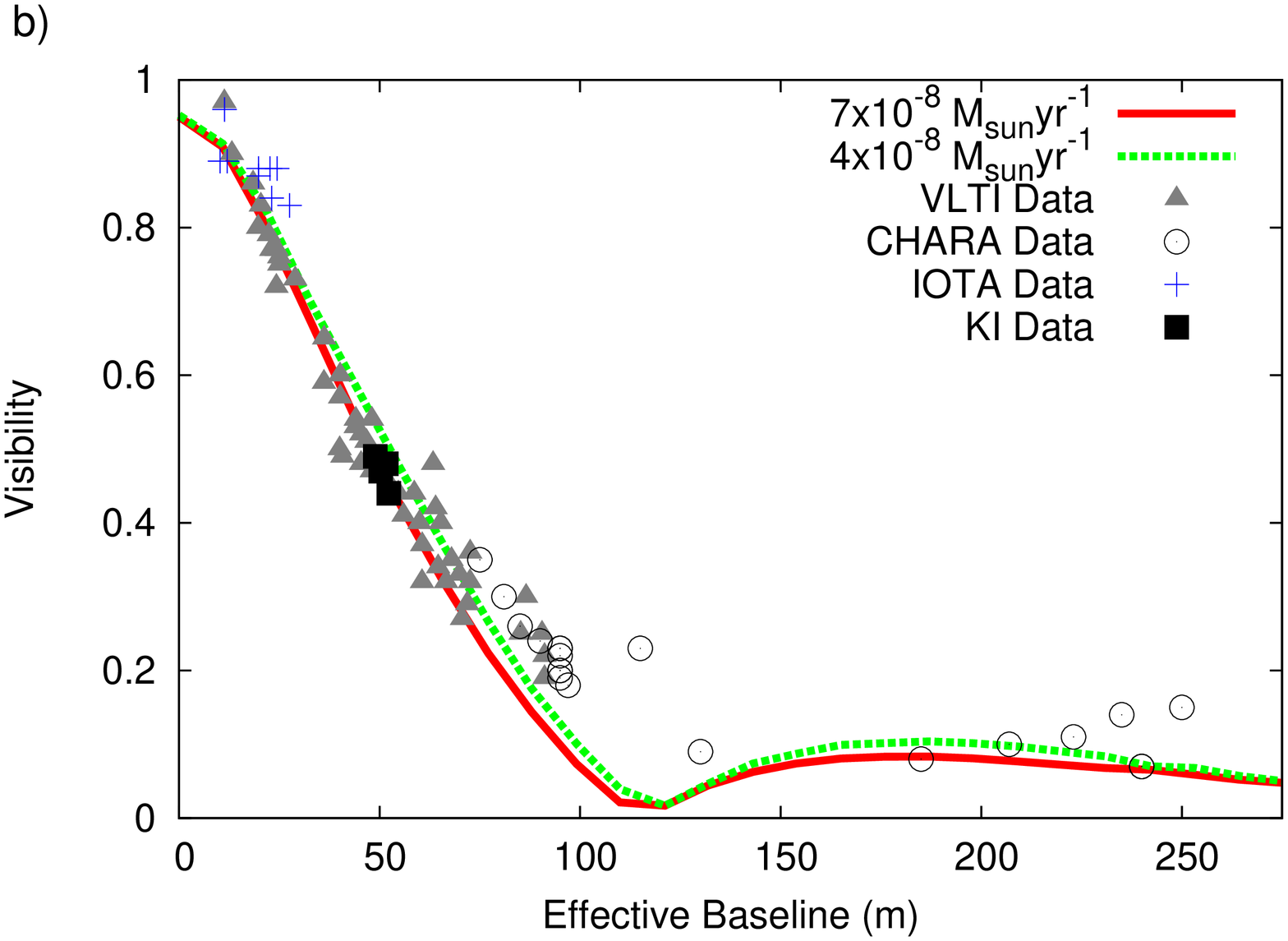} 
\end{center} 
\caption{Model NIR visibilities for AB Auriga (a) and MWC~275 (b), corresponding to the ``best fit'' parameter combinations presented in Figures~\ref{fig3}a and~\ref{fig3}b, respectively. The VLTI data points are from \citet{BenistyEtc10}, whereas the references for the other data sets are given in \citet{TannirkulamEtc08b}.} 
\label{fig4}
\end{figure*}

\subsection{Synthetic Visibilities}
\label{subsec:visibility}

As was already remarked in Section~\ref{subsec:SED}, the ability to fit the SED is not sufficient for distinguishing between competing interpretations of the NIR excess. In fact, spherical dusty halo models \citep[e.g.,][]{VinkovicEtc06} can also account for the NIR bump and, as pointed out by DM10, may even be more consistent than the ``puffed-up inner rim'' model with the apparent lack of a clear correlation between the NIR flux and the disk inclination among the observed 
sources.\footnote{Although the centrifugally driven wind model gives rise to a nonspherical density distribution, the degree of flattening of this distribution is much smaller than that of a thin accretion disk and therefore a strong correlation between the reprocessed flux and the inclination of the symmetry axis is also not expected in this case.} However, with the recent advent  of NIR interferometry it has become possible to start spatially resolving the NIR emission regions of HAe and HBe stars and thereby to better discriminate among the different models. As was noted in Section~\ref{sec:intro}, high-resolution measurements of this type are already available for the two HAe stars modeled in this paper. The existing data for these sources \citep{TannirkulamEtc08b,BenistyEtc10} indicate both that a spherical halo model is very unlikely to reproduce the observations and that (as explained below) a ``puffed-up inner rim'' model is ruled out (unless an additional emission component is present within the sublimation radius). In this subsection we demonstrate that, in contrast, the interferometric results are consistent with the disk-wind model.

While direct imaging is still not technically feasible for the inner regions of protostellar systems, the NIR emission zones of HAe stars are already being probed by long-baseline interferometry. In this technique, the light from two (or more) telescopes is combined to generate an interference patten (``fringes''). The amplitude and phase of these fringes, which depend on the projected distance on the plane of the sky between any given pair of telescopes (the ``baseline''), are combined into a complex quantity, $V$ (the ``visibility''). According to the van Cittert--Zernike theorem, the visibility represents a Fourier transform of the spatial intensity distribution,
\begin{equation} 
\label{eq:visibility}
V(u,v)= \int \int I(\alpha,\beta) e^{2\pi i(\alpha u+\beta v)} d\alpha \, d\beta \, ,
\end{equation}
where $I(\alpha, \beta)$ is the intensity distribution on the sky as a function of the angular coordinates (direction cosines) $\alpha$ and $ \beta$. The spatial frequencies $u$ and $v$ are the normalized (by the observation wavelength $\lambda$) projections of the baseline on a plane perpendicular to the line of sight: $ u=B_x/\lambda$ and $ v=B_y/\lambda$. It should be evident from the form of Equation~(\ref{eq:visibility}) that an extended object with a Gaussian-like intensity profile would produce a visibility that is also Gaussian-like --- smooth and single-peaked. By contrast, the Fourier transform of the intensity profile of a 
uniform-brightness ring exhibits the opposite behavior, displaying large secondary peaks. Thus, the presence or absence of a secondary bounce in the interferometric visibility curve indicates whether the emission region is ``sharp'' (like the uniform ring) or ``fuzzy'' (like a Gaussian). For a review of visibilities and of how to interpret them, see \citeauthor{Berger03} (\citeyear{Berger03}; see also Figure~5 in DM10). 

The basic rim model, by construction, involves a very ``sharp'' emission region and therefore implies a pronounced secondary bounce in the visibility curve. As we already noted in Section~\ref{sec:intro}, such a bounce is clearly absent in the measured visibilities of AB Aur and MWC~275, which are found instead to be relatively flat beyond their initial dropoff. We now show that the smooth, extended dust distribution that characterizes the hydromagnetic disk-wind model naturally accounts for the shape of the visibilities measured in these sources. The predicted visibilities are derived by carrying out the integral in Equation~(\ref{eq:visibility}) using the intensity distribution obtained from the MCRT calculation. We consider a wavelength interval that is centered on $2.2\, \micron$ (corresponding to a typical K band) and plot all visibilities as a function of the  ``effective''  baseline
\begin{equation} 
\label{eq:B_eff}
B_{\rm eff}\equiv B_{\rm projected} [\cos^2{\delta} + \cos^2{\epsilon} \, \sin^2{\delta}]^{1/2}\, ,
\end{equation}
where $\delta $ is the angle between the baseline direction and the major axis of the (projected) disk, and $\epsilon$ is the angle between the disk's rotation axis and the line of sight. As pointed out by \citet{TannirkulamEtc08a,TannirkulamEtc08b}, the quantity in brackets properly accounts, in the case of a flat disk, for the dependence of the interferometric resolution on the disk's inclination and position angle.

Figure~\ref{fig4} shows the visibility curves in the disk-wind model that correspond to the ``best fit'' parameters obtained from the SED modeling in Section~\ref{subsec:SED}: Figure~\ref{fig4}a is for AB Aur and corresponds to the best-fit spectra presented in Figure~\ref{fig3}a, whereas Figure~\ref{fig4}b is for MWC~275 and corresponds to Figure~\ref{fig3}b. In obtaining these curves, we used values of the disk inclination angle $\epsilon$ of $21^\circ$ and $48^\circ$ for AB Aur and MWC~275, respectively \citep[see][]{TannirkulamEtc08b}. Considering first AB Aur, the two curves shown in Figure~\ref{fig3}a correspond to distinct combinations of the parameters $\dot M_{\rm out}$ and $T_{\rm sub}$ but give, as we already noted in Section~\ref{subsec:SED}, similarly good fits to the SED data for this source.  Figure~\ref{fig4}a reveals that the visibility data can be used to break this degeneracy: It is seen that the $T_{\rm sub} = 1850\,$K, $\dot M_{\rm out}=4\times 10^{-8}\, M_\odot\, {\rm yr}^{-1}$ model both matches the short- and medium-baseline data very well \emph{and} provides a clearly superior fit to the lower-$T_{\rm sub}$, higher-$\dot M_{\rm out}$ model. Figures~\ref{fig3}a and~\ref{fig4}a thus illustrate how the simultaneous modeling of SED and visibility data can be used to narrow down the choice of ``best fitting'' model parameters. These figures also demonstrate that, in contradistinction to the rim model, the dusty disk-wind model can by itself account for both the SED and the short/medium-baseline visibility data for this bright HAe star. However, our model fits underestimate the long-baseline ($B_{\rm eff} > 300$) visibility data for this source. It is quite plausible (in view of the expected multiple visibility peaks) that the inclusion of a (subordinate) inner-rim emission component could improve the fit for this spatial-frequency regime, but this remains  to be demonstrated by an explicit calculation.

Unlike the situation depicted in Figure~\ref{fig4}a, the visibility curves shown in Figure~\ref{fig4}b for MWC~275, which correspond to the two parameter combinations used in Figure~\ref{fig3}b, are barely distinguishable. We recall that these two models differ only in the magnitude of $\dot M_{\rm out}$, with the two listed values chosen to account for the observed flux variability in this source. The visibility data shown in the figure were collected over different periods and do not directly correspond to the SED data presented in Figure~\ref{fig3}b. The fact that the predicted visibilities are nearly identical is consistent with the apparent concordance of the different data sets in the low-$B_{\rm eff}$ regime where they overlap. It is conceivable, however, that future multi-baseline, high-time-resolution interferometric observations might be able to help identify the physical mechanism (e.g., the ejection of dusty clouds; see Section~\ref{sec:discuss}) that is responsible for the NIR variability in this class of sources.

Another useful quantity that can be measured interferometrically when at least three telescopes are involved in the observation 
is the ``closure phase'' --- the sum of the fringe phases for three baselines that form a triangle. The value of this quantity serves as a probe of the degree of point symmetry of the source: The closure phase vanishes for a centrosymmetric image, and it increases the more the emission is concentrated off center (or ``skewed''). \citet{MonnierEtc06} carried out a NIR closure-phase survey of a sample of 14 HAeBe stars (including AB Aur and MWC~275) and found that only six (including AB Aur) exhibited statistically significant (although generally small) values. These results are at odds with the original version of the rim model, in which the puffed-up edge of the disk was assumed  to be perfectly vertical. Since only the side of the rim that faces the star emits in this picture, this model predicts NIR images that are strongly skewed for nonzero inclination angles. The discrepancy is mitigated in the subsequently developed ``rounded rim'' models (e.g., \citealt{IsellaNatta05,TannirkulamEtc07}; see Figure~4 in DM10 for a pictorial representation). The overall appearance of the NIR images that are obtained in the dusty disk-wind model  closely resembles the basic structure of the images that are derived using the rounded-rim model. However, given the more extended nature of the emission region in the wind model, it may be expected that the NIR images obtained in this case would be more centrosymmetric than those predicted by the rounded-rim model.

\subsection{Contribution of the Region Interior to $r_{\rm sub}$}
\label{subsec:interior}

As discussed in Section~\ref{subsec:setup} (see Figure~\ref{fig1}), the wind launching region in our model extends down to a radius $r_{0\rm min}$ that is generally smaller than $r_{\rm sub}$. As we already noted, this picture is supported by 
Br$\gamma$ line observations \citep[e.g.,][]{TatulliEtc07,KrausEtc08}. The disk truncation radius, which may be determined by the interaction of the accretion flow with the stellar magnetosphere (see Equation~(\ref{eq:r_m})), could be even smaller. But even though the outflow and  the disk at $r<r_{\rm sub}$ are dust free  (unless they contain highly refractory grains), they can still contribute to the NIR emission. We now estimate this contribution.

Since the interior outflow does not contain dust, it remains optically thin to the bulk of the stellar radiation. It can, however, be heated by internal processes as well as by ionizing radiation from the central star. \citet{Safier93a} examined the thermal structure of disk-driven hydromagnetic winds in cTT stars and identified a feedback mechanism involving ambipolar diffusion (or ion--neutral drag) in the initially weakly ionized disk material that feeds the outflow. He showed that, for a given dynamical structure determined from a solution of the wind equations (which is obtained under the usually well-justified assumption that thermal forces do not strongly influence the flow dynamics), the ambipolar diffusion heating rate scales inversely with 
$\gamma \rho_{\rm i}$, where $\gamma$ is the ion--neutral collisional coupling coefficient and $\rho_{\rm i}$ is the ion mass density. This shows that, so long as $\rho_{\rm i}$ is small, the heating could be strong. \citet{Safier93a} found that this process indeed leads to a rapid increase of the temperature to $\sim 10^4\,$K (see his Figures~5 and~6): At this value the collisional ionization of hydrogen starts to  measurably increase $\rho_{\rm i}$, which in turn causes the heating rate to decline. This mechanism is evidently quite robust, although its effectiveness could be reduced if an external ionization source is present. We expect the same process  to operate also in the dust-free regions of HAe disk outflows, and in Figure~\ref{fig6} we present a schematic of the resulting thermal structure based on Safier's (\citeyear{Safier93a}) calculations. The wind can be roughly divided into three regions: (1) a cool ($\sim 1500\,$K) molecular zone that occupies a thin layer at the bottom of the outflow near the surface of the disk;  (2) a  warmer ($\sim 3000\,$K) atomic and molecular zone that occupies a slightly wider layer directly above the cool zone; and (3) a hot ($\sim 10^4\,$K) low-density region that lies above the previous two and extends to the edge of the modeled domain. In Safier's (\citeyear{Safier93a}) calculations, the temperature of the upper layer was invariably $\gtrsim 10000\,$K and it was characterized by a high degree of ionization. However, subsequent calculations by \citet{ShangEtc02} determined that, when the temperature is $\sim 10000\,$K and the gas is fully atomic, the coupling coefficient $\gamma$ is
about an order of magnitude higher than the ``cold'' value used in \citet{Safier93a}, which would lead to a commensurate  reduction in the ambipolar diffusion heating rate. But Shang et al.'s (\citeyear{ShangEtc02}) results also indicate that, so long as the gas is mostly molecular, Safier's (\citeyear{Safier93a}) adopted value for $\gamma$ should roughly apply (see their Equation~(E8)). Given that molecular hydrogen is expected to survive in the flow until the temperature reaches $\sim 5000\,$K, 
at which point it rapidly dissociates (see Section~4.6 in \citealt{Safier93a}), we estimate that the temperature of the uppermost wind region would level off at that value in the absence of any other heating mechanism. \citet{ShangEtc02} suggested that one conceivable mechanism for additional heating is the dissipation of fluctuations that are induced at the source.
They wrote down a parametrized phenomenological expression for the rate of such mechanical heating, and \citet{ObrienEtc03} showed that, with the inclusion of this additional mechanism, a disk wind's temperature could in principle again reach (or even surpass) $10000\,$K. For completeness, we thus consider the following two options for the temperature $T_3$ of Region 3: $T_3=10000\,$K, for which we label the zone as `3a', and $T_3=5000\,$K, for which we use the label `3b'. Note that, when mechanical heating is strong enough to raise the wind temperature to $10000\,$K, it may largely eliminate the cooler underlying layers (see Figure~1 in \citealt{ObrienEtc03}). However, as we show below, this modification would have little effect on the NIR appearance of the source.
 
\begin{figure}
\centering
 \includegraphics[ width=0.5\textwidth, angle=0]{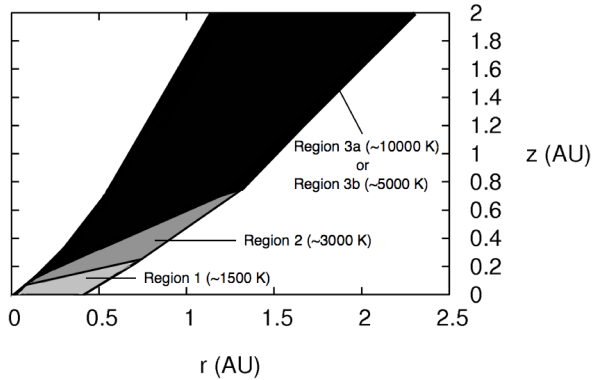}
\caption{Schematic of the thermal structure of an internally heated wind. The two options for the temperature $T_3$ of Region~3 are discussed in the text. If the wind were heated instead by the star's ionizing radiation, its structure would be similar to that of Region~3a.} 
\label{fig6}
\end{figure}

HAeBe stars emit Lyman continuum photons (at a rate $Q_*$ in the range $\sim 10^{43}-10^{45}\, {\rm s}^{-1}$; e.g., \citealt{AlexanderEtc05}), which can also contribute to the heating of the wind. We can utilize the approximation of a point source and the self-similarity (in the spherical radial coordinate) of the \citet{BP82} wind solution to express the outer spherical radius $R_{\rm out}$ of the ionized zone within the wind (the ``Str\"omgren region'') as a function of the polar angle $\theta$,
\begin{equation}
\label{eq:R_max}
R_{\rm max} = R_{\rm in} \exp\{Q_*/[4\pi \alpha R_{\rm in}^3 n^2(R_{\rm in})]\}\, ,
\end{equation}
where $R_{\rm in}$ (which is also a function of $\theta$) is the spherical radius of the intersection point of a ray from the center in the direction $\theta$ with the innermost streamline of the wind, $n(R_{\rm in})$ is the wind's particle density at that location, and $\alpha$ is the hydrogen recombination coefficient; we assume that the gas is already atomic in the region of interest. Using this expression, we find that only a thin layer in the vicinity of the innermost streamline is photoionized for the values of $\dot M_{\rm out}$ implied by the spectral fitting described in Sections~\ref{subsec:SED} and~\ref{subsec:visibility}. We therefore do not consider this case further in this Section. We note, however, that when $\dot M_{\rm out}$ is low enough for the ionizing photons to penetrate the wind, the shape and temperature of the Str\"omgren region closely resemble those of Region~3a in 
Figure~\ref{fig6}.

In calculating the NIR emission from the optically thin, dust-free wind component, we used average Planck opacities 
that were provided to us by J. Ferguson (private communication 2012) since the relevant region in the temperature--density parameter space is not included in previously published opacity tables \citep[e.g.,][]{FergusonEtc05}. We treated the underlying disk component as a flat slab that intercepts the radiation from the star (characterized by an effective temperature $T_*$ and radius $R_*$) and reradiates it as a blackbody of temperature
\begin{equation}
\label{eq:T_in}
T(r) =\left(\frac{1}{8}\right )^{1/4} \left (\frac{R_*}{r}\right )^{1/2} T_*  \, 
\end{equation}

\citep[e.g.,][]{Hubeny90}. The inner disk should be optically thick for the mass accretion rates inferred in our model, and while this poses a problem for the inner rim scenario \citep[e.g.,][]{MuzerolleEtc04}, it has no adverse effect on the disk outflow interpretation of the NIR bump, in which a dusty wind intercepts the stellar photons well above the disk surface. The results we show were obtained under the assumption that the disk's emission region coincides with the wind launching region (with inner radius $r_{0\rm min} = 0.05\,$AU). Figure~\ref{fig7} shows the contribution of the interior emission components to the SED of a source like AB Aur for our two representative choices of the temperature $T_3$ in Region~3 of the wind. It is seen that the disk emission dominates, but that there is also a measurable contribution at the shortest wavelengths from wind Region~3, particularly in the $T_3=10000\,$K case. The reason why the main contribution from the wind comes from Region~3a is that the emitted flux is 
$\propto \rho V T^4$,  and this region has the highest temperature and comprises by far the largest volume $V$, which more than make up for its lower density $\rho$. For the parameter range that we explored, the interior components contribute at a level of only $\sim 10-20\, \%$ of the dusty wind's emission. However, as Figure~\ref{fig7} demonstrates, this contribution suffices to reduce the required wind outflow rate by nearly a factor of 2 compared to the fit shown in Figure~\ref{fig3}a for the same wind model (C) and subilmation temperature ($1500\,$K).\footnote{We do not use our best-fit model for AB Aur, which is characterized by $T_{\rm sub} = 1850\,$K, to illustrate the contribution of the interior wind and disk components to the SED because this contribution becomes hard to discern in the corresponding figure for the higher-$T_{\rm sub}$ case.}
 
Figure~\ref{fig8} shows the effects of the interior emission components depicted in the left panel of Figure~\ref{fig7}
on the predicted K-band visibility. It is seen that, overall, these effects are minor, which is consistent with the relatively low contribution of these components to the total flux at these wavelengths. The qualitative changes to the visibility curve can be understood from the fact that the interior components contribute emission at smaller radii: This both reduces the effective size of the emission region somewhat, which shifts the first minimum to a slightly longer baseline, and further ``smears out'' the emission, which is reflected in a small drop in the amplitudes of the secondary peaks. The inference that changing the outflow rate by nearly a factor of 2 in our homogeneous wind model does not have a major impact on the visibility curve is consistent with the conclusion we have already drawn from modeling the visibilities for the ``high'' and ``low'' states of MWC~275 (see Figure~\ref{fig4}b). The fact that the addition of the interior components does not modify the predicted visibility in a significant way implies that our best-fit model for AB Aur --- which is reproduced here from Figure~\ref{fig4}a --- still requires a comparatively high dust sublimation temperature.\footnote{This result is similar to that obtained by \citet{TannirkulamEtc08b} in the context of the rim model. However, in that paper they also needed to invoke a major interior emission component to account for the full strength of the NIR bump and for the lack of a pronounced secondary bounce in the visibility.}

\begin{figure*}
\centering
 \includegraphics[ totalheight=0.7\textheight, angle=270]{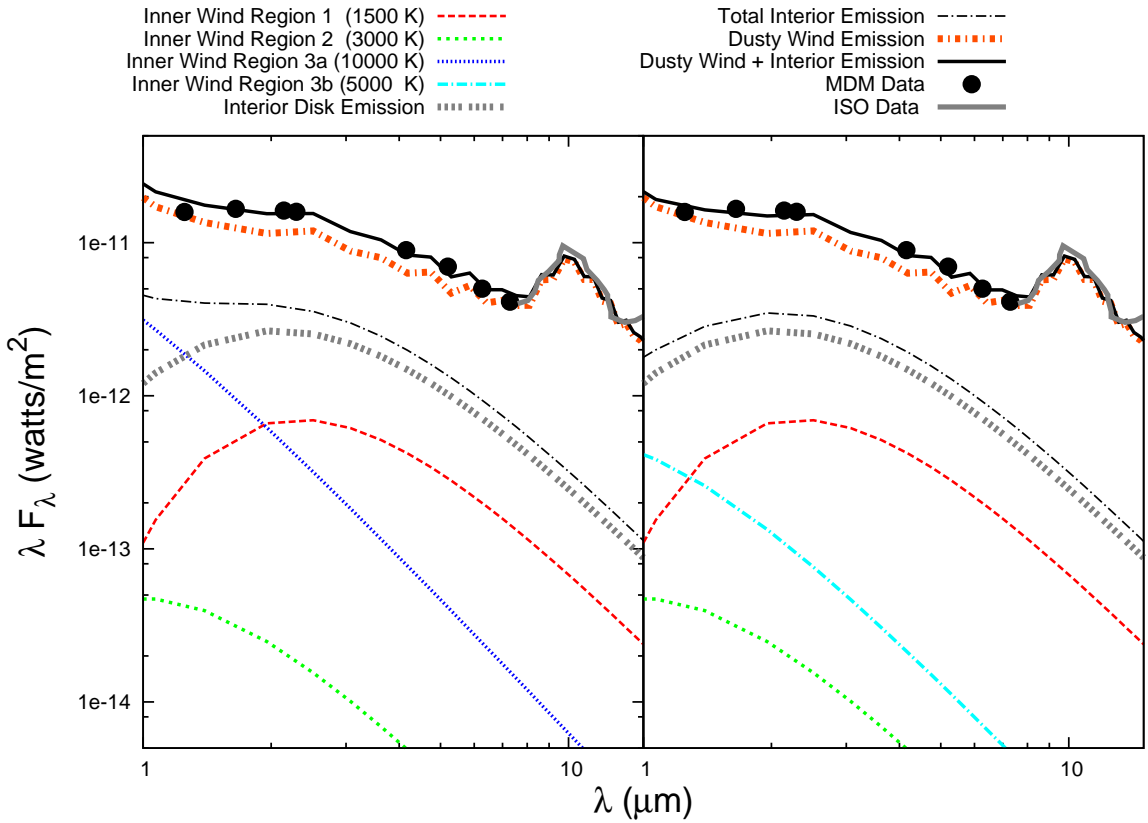}
\caption{Contribution of the interior wind and disk to the model SED fit for AB Auriga. The contribution of the three wind regions described in Figure~\ref{fig6} as well as of the disk component are shown for the two adopted representative temperatures for wind Region~3: $T_3=10000\,$K (Left) and $T_3=5000\,$K (Right). Also shown are the $T_{\rm sub}=1500\,$K, $\dot M_{\rm out}=4\times 10^{-8}\, M_\odot\, {\rm yr}^{-1}$ dusty wind model from panel (d) of Figure~\ref{fig2} and the data points displayed in that figure.}
 \label{fig7}
\end{figure*}

\begin{figure*}
\centering 
 \includegraphics[ totalheight=0.5\textheight, angle=270]{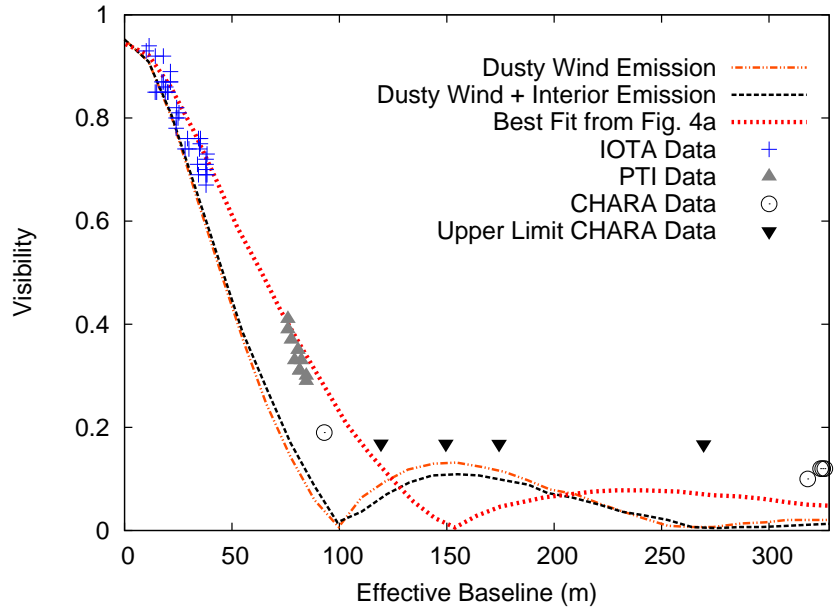}
\caption{Contribution of the interior wind and disk to the model NIR visibility fit for AB Auriga. The dusty wind model and the interior wind and disk components are those shown in the left panel of Figure~\ref{fig7}. The data points are the same as those displayed in Figure~\ref{fig4}a. Also shown for reference is the best-fit model from Figure~\ref{fig4}a, corresponding to a sublimation temperature of $1850\,$K and an outflow rate of $4 \times 10^{-8}  M_\odot\, {\rm yr}^{-1}$. }
  \label{fig8}
\end{figure*}

\section{Discussion}
\label{sec:discuss}

The availability of both SED and high-resolution visibility data for several bright HAe stars has made it possible to place strong constraints on physical models of the NIR excess emission in these sources. The observations are consistent with reprocessing of the stellar radiation by a nonspherical distribution of dust, and they have been commonly interpreted in terms of a disk with a puffed-up inner rim. The flat shape of the spectrum in the $\sim 3\, \micron$ bump regime points to comparatively large ($\gtrsim 1\, \micron$) grains that have a near-gray opacity, and the measured drop-off of the visibility curve at short baselines indicates a size of the inner rim that, when identified with the dust sublimation radius, yields the approximate value of the dust sublimation temperature. However, detailed models of the rim have indicated that this component by itself cannot account for the strongest observed NIR bumps. A second, and perhaps even more serious, difficulty with this interpretation is the apparent absence of a prominent secondary bounce in the visibility curve at intermediate baselines, which should be present if the emission region has a sharp edge. To address these two problems, several authors have invoked the presence of an extended NIR-emitting disk component of temperature $T>T_{\rm sub}$ within the dust sublimation radius. In the case of MWC~275, for example, \citet{TannirkulamEtc08a,TannirkulamEtc08b} suggested that this component most likely represents hot gas, but \citet{BenistyEtc10} argued that a realistic gaseous disk would not be consistent with the spectral and interferometric constraints on this source and proposed that the hot inner disk component consists instead of refractory dust. However, the implied sublimation temperature of this dust is significantly higher than what is commonly assumed.

As we demonstrated in Section~\ref{sec:results}, a hydromagnetic disk wind, which produces a stratified density distribution of dusty gas beyond the dust sublimation radius, naturally accounts for both the spectral and the interferometric properties of the observed HAe systems. In particular, for plausible outflow parameters, the optically thick (to the stellar radiation) region of such a wind subtends a much larger solid angle at the star than the puffed-up inner rim of the disk, making it possible for this model to account for even the strongest measured NIR excesses. In addition, given that the NIR-emitting region in the wind is generally spatially extended, the predicted visibility curves lack a pronounced secondary bounce.
 These features are in accord with the observations and contrast with the predictions of the spatially localized rim scenario. In the context of our simplified model, similar spectral fits to the NIR bump can be obtained by using different combinations of the parameters $\dot M_{\rm out}$ (the wind mass outflow rate) and $T_{\rm sub}$ (the nominal dust sublimation temperature). Using the example of AB Aur, we showed that this degeneracy can be broken by also fitting the visibility curve. Although in this paper we focused on the application of this model to HAe stars, it should be relevant also to other effectively low-$L_*$ protostars, including, in particular, cTT stars. 

The hydromagnetic disk-wind model also provides a natural framework for interpreting the pronounced NIR variability exhibited by sources such as MWC~275. As we showed in Figure~\ref{fig3}b, the NIR burst measured  by \citet{SitkoEtc08} can be attributed to an increase in the mass outflow rate (and hence in the density and optical depth of the wind) by $\sim 75\, \%$. In the simple self-similar wind model that we have adopted, this increase could be brought about by a commensurate increase in the mass inflow rate. However, in general, the disk's capacity to launch a centrifugally driven outflow from any given radius is determined by several factors, which include the magnetic field-line inclination at the disk surface (see Section~\ref{subsec:CDW}) as well as the degree of ionization and the magnetic-to-thermal pressure ratio inside the disk \citep[e.g.,][]{SalmeronEtc07}. Each of these factors is affected by different physical processes that may act on different spatial scales, and it is therefore also possible that there could be pronounced changes in the local outflow rate that are not linearly correlated with the local behavior of $\dot M_{\rm in}$. We recall in this connection our argument that the launching zone of the dusty outflow is effectively self-limiting to the narrow region between $r_{\rm sub}$ and $\sim 2\, r_{\rm sub}$ (see Section~\ref{subsec:setup} and Appendix~\ref{sec:AppA}). It can therefore be expected that any changes in the wind-launching conditions in this radially localized region of the disk would have a direct impact on the magnitude of the generated NIR excess.

Several protostars, including MWC~275 \citep{WisniewskiEtc08}, have been found to exhibit photometric variability in the outer regions of the circumstellar disk on time scales (of a few years) that are much shorter than the local dynamical time. This behavior is most readily interpreted in terms of variable shadowing by density inhomogeneities in the inner disk. \citet{VinkovicJurkic07} pointed out that this shadowing cannot in general be explained in the context of the rim model and advocated an alternative explanation in terms of dusty ``clouds'' that are ejected from the disk and transit across the line of sight to the source. The same clouds can also produce the observed NIR variability and, in a similar vein, may account for the anti-correlated near- and far-infrared variability that has been observed in the cTT star LRLL~31 \citep{MuzerolleEtc09}. \citet{VinkovicJurkic07} noted, however, that the biggest drawback to this picture was, in their view, ``the lack of a known force capable of lifting a dust cloud out of the disk.'' This difficulty can in principle be overcome within the framework of the disk-wind model, where one can appeal, for example, to the ram pressure of the magnetically driven homogeneous wind component as one possible mechanism for uplifting clouds from the disk.\footnote{\citet{TambovtsevaGrinin08} similarly suggested that inhomogeneities in the inner disk wind, which they attributed to a generic turbulence of the outflowing gas, could produce variable shadowing effects in the outer disk.} This process was previously considered in connection with the interpretation of maser disks in AGNs (\citealt{KartjeEtc99}; see also \citealt{KondratkoEtc05}). As we noted in Section~\ref{sec:intro}, the dusty disk-wind interpretation of the $\sim 3\,\micron$ bump was also originally proposed in the context of AGNs. It is interesting to observe in this connection that detailed studies of the infrared emission properties of this class of sources have provided strong evidence that typical disk outflows in AGNs are, in fact, clumpy \citep[e.g.,][]{ElitzurShlosman06,NenkovaEtc08}.

To check on the self-consistency of the scenario wherein a cloud is uplifted by the ram pressure of a disk wind, we estimate the maximum hydrogen column density of an uplifted cloud (subscript `c'), $N_{\rm c,max} \equiv n_{\rm c} R_{\rm c}$ (where $n_{\rm c}$ is the cloud's mean particle density and $R_{\rm c}$ is its radius), by balancing the upward wind ram-pressure force on the cloud against the downward tidal gravitational  force. This yields $N_{\rm c,max} \approx \dot M_{\rm out}/[(GM_*r)^{1/2}\mu m_{\rm. p}] = 2.3\times 10^{22}\, (\dot M_{\rm out}/5\times 10^{-8}\, M_\odot\, {\rm yr}^{-1})(M_*/2\, M_\odot)^{-1/2}(r/0.3\, {\rm AU})^{-1/2}\; {\rm cm}^{-2}$ (where $\mu \approx 2.33$ is the molecular weight and $m_{\rm p}$ is the mass of a hydrogen nucleus; cf. Equation~(24) in \citealt{KartjeEtc99}). A cloud with this column will indeed block most of the stellar radiation that it intercepts. We can also roughly estimate the transit time $t_{\rm tr}$ of a cloud of this type that moves across the line of sight in the vicinity of the disk surface. Taking a typical cloud density $\rho_{\rm c}$ and projected speed $v_{\rm tr}$ to be $\sim \rho(r=r_{\rm sub},z=0)$ and $\sim \psi_0 v_{\rm K}$, respectively (see Equations~(\ref{eq:rho}) and~(\ref{eq:rho1})), we obtain $t_{\rm tr}\equiv 2R_{\rm c}/v_{\rm tr}  \approx 1.2\,$yr for our fiducial parameter values. This estimate is consistent with the inferred values in a source like MWC~275 \citep{SitkoEtc08,WisniewskiEtc08}. We note, however, that alternative mechanisms for generating a clumpy outflow from a magnetized disk --- for example, a process that resembles coronal mass ejections in the Sun --- are also conceivable.

\section{conclusion}
\label{sec:conclude}

In this paper we present a new interpretation of the near-infrared excess, dominated by a $\sim 3\, \micron$ ``bump,''  that is exhibited by the monochromatic luminosities of protostellar systems (notably low-mass classical T Tauri stars and intermediate-mass Herbig Ae and Be stars). 
The bump can be approximated by a thermal emission component whose temperature ($\sim 1500\,$K) is of the order of the sublimation temperature of typical interstellar grains, which has motivated the suggestion that it is produced in the vicinity of the dust sublimation radius in these systems. This interpretation is supported by the finding that the interferometrically measured sizes of low-luminosity sources scale with the bolometric luminosity as expected for $r_{\rm sub}$. The most widely invoked scenario until now has been the ``puffed-up inner rim'' model, which postulates that the protostellar accretion disk in sources with pronounced NIR excess is optically thin to the stellar photons, so that the latter can propagate  along the equatorial plane until they are absorbed by the dust in the disk at $r_{\rm sub}$ and subsequently reemitted at NIR wavelengths. In this picture, the inner edge of the dusty disk expands as a result of heating by the absorbed radiation, which increases the fraction of stellar photons that are intercepted and reprocessed by the disk. Although this model is able to  relate the site of the bump emission to the dust sublimation radius, it has faced two important difficulties: It cannot account for the observed strength of the bump in the brightest sources, and it predicts (on account of the inherent sharpness of the dusty disk's inner edge) prominent secondary peaks in the interferometric visibility curve, which are not observed. 

According to our proposed interpretation, the dominant contribution to the NIR bump in protostars comes from the reprocessing of the stellar radiation in a dusty hydromagnetic disk wind. In this picture, the dust is uplifted from the disk by a centrifugally driven wind, and since dust is present only beyond $r_{\rm sub}$, this naturally accounts for the association of the bump emission region with that radius. Since most of the stellar photons in this scenario are intercepted above the disk surface, it is not necessary, as in the rim model, for the disk to be optically thin to the stellar radiation at $r<r_{\rm sub}$ . Furthermore, since the dusty wind subtends a much larger solid angle at the star than the dusty disk's inner rim, this model can reproduce the strength of the bump even in bright sources such as the HAe stars AB Auriga and MWC~275. The lack of a pronounced secondary bounce in the visibilities measured in these sources by high-resolution interferometry is also naturally explained by the extended nature of the emission region in the disk-wind model. In fact, using a simplified wind/disk model and a specially constructed Monte Carlo radiative transfer code, we showed that a single combination of physically plausible wind parameters can generate a good fit to both the SED and the visibility for each of the above two sources. In the case of AB Aur, we  demonstrated how the parameter degeneracy that arises when one fits the SED alone can be lifted when both the SED and the visibility are fitted simultaneously. 

A centrifugally driven wind along a large-scale, ordered magnetic field is a leading candidate for the physical mechanism that underlies the ubiquitous energetic outflows in protostellar systems. There are observational indications that at least some of the outflows are launched from the circumstellar accretion disk on scales that are comparable to $r_{\rm sub}$, and that some outflows contain dust that likely originates in the disk. Based on the wind parameters of our model spectral fits, we argued that the spatial extent of the launching region of a dusty disk outflow could be self-limiting to a narrow radial range (between 
$r_{\rm sub}$ and $\sim 2\, r_{\rm sub}$) on account of the shielding by the uplifted dust of stellar FUV photons that play a key role (through photoevaporation) in the mass loading of the wind. We also suggested that the disk outflow model provides a natural framework for interpreting the spectral variability measured in several of these sources. In the case of MWC~275, we showed that the reported recent $\sim 30\,\%$ increase in the $1-5\, \micron$ flux could be attributed to a $\sim 75\, \%$ increase in the local mass outflow rate. It has previously been proposed that the NIR variability in protostars is associated with the ejection of dusty clouds, and we pointed out that the ram pressure force exerted by a homogeneous hydromagnetic wind is one plausible means of uplifting such clouds from the disk. 

It is interesting to note that a $\sim 3\, \micron$ NIR bump is also observed in certain AGNs, and that the association between the size of the region from which this emission originates and the dust sublimation radius was first made in that context. The interpretation discussed in this paper, wherein the NIR excess is attributed to the reprocessing of radiation from a central continuum source by a dusty disk outflow, was also previously considered in connection with Seyfert galaxies and quasars. 
More recent modeling of the infrared emission in AGNs has established that the dusty winds in these sources are likely clumped, which brings out yet another possible similarity with protostellar disk outflows.

There is evidence that low-luminosity low- and intermediate mass protostars (comprising classical T Tauri stars, Herbig Ae stars, and low-$L_*$ Herbig Be stars) differ from high-luminosity Herbig Be stars in their observational manifestations, including the 
H$\alpha$ line polarization \citep{VinkEtc05}, the location on the size--luminosity diagram \citep{MonnierEtc05}, and the distribution of visibilities \citep{VinkovicJurkic07}. In the context of the dusty disk-wind scenario, high-luminosity sources are distinguished by the fact that the effect of radiation pressure on grains may affect the dynamical structure, and consequently the radiative properties, of the dusty portions of their disk outflows \citep[e.g.,][]{KK94}. We plan to investigate the extent to which this effect may influence the observed characteristics of high-$L_*$ protostars in future work.

\acknowledgments
We thank Jason Ferguson for kindly providing us with the opacity tables for the calculation described in 
Section~\ref{subsec:interior}. We also acknowledge useful suggestions by the referee that helped improve the presentation. This research was supported in part by NSF grant AST-0908184 as well as by a NASA Earth and Space Science Fellowship and a Brinson Foundation University of Chicago Predoctoral Fellowship awarded to A.B.
 
\appendix
\section{Disk Shielding by Dusty Wind}
\label{sec:AppA}
\begin{figure*}
\begin{center}
\includegraphics[totalheight=.5\textheight, angle=270]{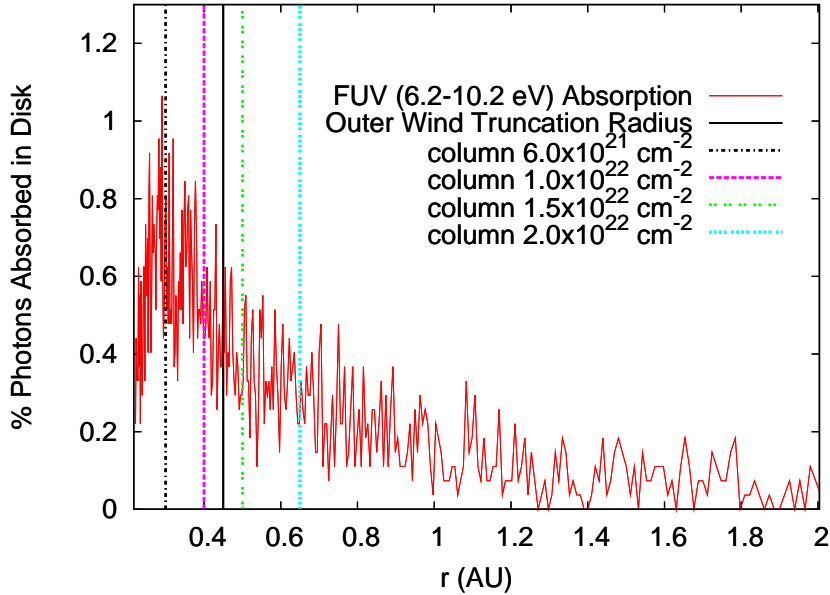}
\end{center}
\caption{Spatial distribution of the stellar FUV photons that reach the disk surface, derived from a MCRT calculation for the best-fitting model of AB Auriga (corresponding to $\dot M_{\rm out} = 4\times 10^{-8}\, M_\odot\, {\rm yr}^{-1}$ and $T_{\rm sub} = 1850\,$K; see Figures~\ref{fig3}a and~\ref{fig4}a). The solid vertical line indicates the location of the adopted outer truncation radius of the disk wind, whereas the other vertical lines indicate the locations on the disk surface where a column density through the dusty, molecular wind to a point at a latitude of $60^\circ$ on the surface of AB Aur would have the value marked in the figure.}
\label{figA1}
\end{figure*}

In Section~\ref{subsec:setup} we noted that, if external heating of the disk surface is required for efficient mass loading of the wind \citep[cf.][]{CasseFerreira00,PesentiEtc04}, then a strong dusty outflow may shield the outer regions of the disk from the heating stellar X-ray and FUV radiation and could thus be self-limiting in its radial extent. For the wind and stellar parameters of the HAe systems considered in this paper, we suggested that the dusty outflow, which starts at the dust sublimation radius $r_{\rm sub}$, would typically not extend much beyond $\sim 2\, r_{\rm sub}$. Since this effect could in principle apply to other types of protostars and conceivably also to other astrophysical systems --- such as AGNs ---  in which a disk wind, especially one that is dusty, could be self-shielding in this way, it may have important implications to our general understanding of astrophysical disk winds. We therefore consider this effect here in somewhat greater detail.

The basic figure of merit that we identified is the column density to the absorption of FUV ($6-13.6\,$eV) photons, which, as \citet{GortiHollenbach09}, for example, pointed out, depends in detail on the opacity of the uplifted dust but is typically $\sim 10^{22}\, {\rm cm}^{-2}$. Incidentally, the absorption of X-ray and FUV photons is considered in the just-cited reference in connection with the photoevaporation of circumstellar disks by the stellar radiation. This may, in fact, be a useful way of thinking of the effect of these photons on launching the disk outflow. As explained in \citet{GortiHollenbach09}, the evaporative outflow that is initiated in this way would not be strong enough to overcome the downward pull of tidal gravity in the inner disk region that is of interest to us. We note, however, that the initiation of this outflow may nevertheless have an important effect on the mass loading of a hydromagnetic wind, in which magnetic stresses are primarily responsible for overcoming the gravitational force.

A rough estimate of whether the dusty wind is self-shielding to FUV photons can be obtained with the help of the self-similar wind model that we have employed in our calculations. Considering the limiting case of a ray that grazes the disk surface, we can use Equations~(\ref{eq:rho}) and~(\ref{eq:rho1}) to evaluate the hydrogen-nucleus column density of dusty molecular gas beyond $r_{\rm sub}$ to be $N_{\rm H} = (2\, \rho_1 r_1/\mu m_{\rm p})(r_1/r_{\rm sub})^{1/2}$. Using $\mu=2.33$, $r_{\rm sub}/r_1 =0.3$,  and the fiducial parameter values adopted in Equation~(\ref{eq:rho1}), we get  $N_{\rm H}=5.0\times 10^{22}\, {\rm cm}^{-2}$. While this is clearly an upper limit, it is useful in indicating that the bulk of the FUV photons that arrive at the disk from above are likely to be absorbed in the wind over a restricted radial range, so that the wind could be self-limiting in the sense discussed above. We now present the results of explicit calculations that quantify the extent of this radial range.

In a real system, stellar photons can reach the surface of an optically thick disk on account of the finite solid angle that the star subtends at any point on the disk's equator, because of the flaring of the disk's surface, and through scattering by the dust in the wind. However, in our simplified MCRT scheme, in which the star is treated as a point source and the disk is taken to be flat, only dust-scattered photons reach the disk surface. We have verified, however, that the temperature distributions of both the dusty surface layer and the thermalized disk interior that we use to approximate the thermal structure of the dusty disk (see Section~\ref{subsec:spectral}) obey the $T(r)\propto r^{-3/4}$ relation of a flat, optically thick disk that is irradiated by a finite-size star \citep[e.g.,][]{AdamsShu86}. This indicates that the approximation of a point source does not have a significant qualitative effect on the calculated thermal structure of the disk.

Figure~\ref{figA1} shows the results of a MCRT calculation for model parameters appropriate to AB Auriga. The figure shows the spatial distribution of the FUV photons that reach the disk, which are a fraction $\sim 5-10\%$ of the total FUV photons emitted by the star. (For comparison, a flat, infinite, windless disk around a finite-size star intercepts 1/4 of the emitted photon; e.g., \citealt{AdamsShu86}.) The solid vertical line in the figure marks the location ($r_{0\rm max} \approx 2\, r_{\rm sub}$) of the outer wind truncation radius in our model. It is seen that this location roughly corresponds to the FWHM point for the distribution of stellar photons that reach the disk. The other vertical lines were obtained from a complementary calculation, in which the column densities through the dusty, molecular wind to different locations on the disk surface were calculated along rays originating at a latitude of $60^\circ$ (mimicking the possible enhancement of UV emission at the location of a magnetospheric accretion shock on the stellar surface) on a sphere of radius $2.4\, R_\odot$ (the inferred radius of AB Aur) centered at the origin. It is seen that the nominal outer truncation radius lies in between the disk radii that correspond to columns of $1.0\times 10^{22}\, {\rm cm}^{-2}$ and $1.5\times 10^{22}\, {\rm cm}^{-2}$, respectively, which is entirely consistent with the figure of merit for $N_{\rm H}$ adopted above. These results support the idea that the launching zone of a robust dusty outflow in the disks of HAe stars
could be restricted to the vicinity of $r_{\rm sub}$ by the absorption in the dusty outflow of stellar FUV photons that would be needed to mass-load the wind further out.

\section{Monte Carlo Radiative Transfer Scheme}
\label{sec:AppB}

\subsection{Geometry}
\label{subsec:geometry}

As is standard in most MCRT calculations, the determination of events (in this case scatterings and absorptions) occurs in smaller subdivisions or \emph{zones} of the modeled geometry. The zones are taken to be sufficiently small to ensure that the values of variable parameters would typically be nearly constant across them. In the present application, the relevant variables are the grain temperature (calculated in the code) and the gas density (which is an input variable obtained from the wind model). The configuration of zones generally reflects some symmetry in the problem; in the case of the self-similar centrifugally driven winds that we consider, the shape of the streamline (given by Equation~(\ref{eq:R})) is the natural geometry on which the  zone structure can be based.  Since the density of dust in the wind decreases with cylindrical radius $r$, we expect most of the``action'' to be near the inner boundary of the dusty zone at $r_{\rm sub}$, which would call for a higher resolution (smaller zones) in that region. With this in mind, our zones are defined horizontally by streamlines launched 
from logarithmically sampled footpoint radii $r_{0}$. We keep the vertical boundaries of the zones 
uniformly spaced. 

\subsection{Initial Sampling of Source Luminosity}
\label{subsec:sampling}

The MCRT scheme used in this paper is based on the ``packet'' method described in \cite{BjorkmanWood01} and \cite{Lucy99}.  The underlying principle of this method is that photons are sampled from an initial source luminosity $L_{*}$ in the form of $N_{p}$ equal-energy packets.  The energy of each packet is thus given by
\begin{equation} 
E_{p}=\frac{L_{*}\delta  t}{N_{p}} \; . 
\label{eq:B1}
\end{equation}
Packets are strictly monochromatic, implying that, since each packet's energy is held 
constant, the number of photons in a packet is wavelength dependent. This implies that there is a larger number of photons in a packet corresponding to infrared radiation than in a packet representing blue light. In the numerical scheme, the wavelength of any given packet is sampled by assuming that the source is a blackbody of temperature $T_*$. Dividing the emission into $K$ sampled wavelengths, equal-probability wavelengths $\lambda_{K}$ are determined by
\begin{equation}  
\label{eq:B2}
\frac{k+0.5}{K}= \frac{\int_0 ^{\lambda_{k}} B_{\lambda}(T_*)d\lambda}{\int_0 ^{\infty}B_{\lambda}(T_*)d\lambda} \ , \quad\quad k=0,1,2.... K-1 \; ,
\end{equation}
where $B_{\lambda}(T_*)$ is the Planck function.

 A source packet's wavelength is determined in the code by randomly drawing an integer $k$ in the interval $(0,K-1)$ and then using the corresponding $\lambda_{k}$ from Equation~(\ref{eq:B2}). The initial angle $\theta$ at which a packet from the source travels into the system (i.e., through the innermost streamline of the dusty wind region) is taken to be random and is determined from
\begin{equation} 
\label{eq:B3}
\theta=cos^{-1}(2\zeta_{1} -1)  \; .
\end{equation} 
Here, and throughout this Appendix, $\zeta_{i}$ represents a random deviate in the interval $(0,1)$.  
\subsection{Radiative Events}
\label{subsec:events}

As the photon packet travels through the dusty wind, there is a probability for it to interact with the dust grains through either scattering or absorption. Whether or not an event occurs is a function of the absorption and scattering opacities, and the first step in determining whether any event takes place within the zone is to calculate the effective maximum optical depths for these interactions:
\begin{equation}      
\label{eq:C1}
\tau^{\rm a/s}_{\lambda}=\int_{r_{\rm c},z_{\rm c}}^{r_{\rm end},z_{\rm end}}\rho(r,z) 
\kappa_{\lambda}^{\rm a/s} dl
\end{equation}
The superscripts `a' and `s' refer to the absorption and scattering processes, respectively. The point \{$r_{\rm c}$,\,$z_{\rm c}$\} represents the current location (in cylindrical coordinates) 
of the packet, whereas the point \{$r_{\rm end}$,\,$z_{\rm end}$\} represent the location on the boundary of the next adjacent zone that a packet would hit if traveling a maximum possible path length $dl$ through the current zone. The path length $dl$ is a function of both $r_{\rm c}$ and $z_{\rm c}$ and of the packet's current direction angle $\theta$. [Since our scheme is formally 2D, the packet direction has no azimuthal ($\phi$) dependence.] The following 
condition determines whether there will be any event (either scattering or absorption) within the current zone:
\begin{equation} 
\label{eq:C2}
 \zeta_{2} \leq 1- e^{-\tau_{\rm tot}}  \; .  
\end{equation}
The optical depth $\tau_{\rm tot}$ refers to the sum of the absorption and scattering optical depths ($\tau_{\rm tot}=\tau^{\rm a}_{\lambda} + \tau^{\rm s}_{\lambda} $). If this condition does not hold, the packet continues traveling until it reaches \{$r_{\rm end}$,\,$z_{\rm end}$\} and 
the steps outlined in Equations~(\ref{eq:C1}) and~(\ref{eq:C2}) are repeated for the new zone. If the condition~(\ref{eq:C2}) holds and there is an event in the current zone, the location within the zone, \{$r^\prime$,\,$z^\prime$\}, where the event occurs is determined by finding the effective optical depth that the given packet actually reaches, i.e., 
\begin{equation}
\label{eq:C3}
 \zeta_{3}=\frac{ \int^{\{r^\prime,\,z^\prime\}}_{\{r_{\rm c},\,z_{\rm c}\}}\rho(r,z)  \kappa_{\rm tot} dl^\prime }{\tau_{\rm tot}} \; .
 \end{equation}
 Here $\kappa_{\rm tot}$ is the sum of the absorption and scattering opacities. After this distance is determined, a newly drawn random deviate determines what type of event occurs:
 \begin{equation} 
 \label{eq:C4}
  \zeta_{4} \leq \frac{\kappa^{\rm a}}{\kappa^{\rm a} + \kappa^{\rm s}} \ . 
  \end{equation}
When the condition~(\ref{eq:C4}) holds, the event is absorption, whereas if it does not, the event is scattering. 

\subsection{Scattering}
\label{subsec:scattering}

Scattering in this problem is effectively purely elastic, and in this analysis it is also treated as being isotropic. Hence the only variable that is changed by a scattering event is the angle at which the packet is traveling. The new angle after scattering is randomly drawn as in Equation~(\ref{eq:B3}).

\subsection{Absorption}
\label{subsec:absorption}

As the code runs, the number of packets absorbed in each zone, $N_{ij}^{\rm abs}$, is tallied and used to calculate the local grain temperature $T_{ij}$. Since each packet has the same energy, the total energy absorbed in a zone is (using Equation~(\ref{eq:B1}))
\begin{equation} 
\label{eq:E1}
E_{ij}^{abs}=N_{ij}^{abs} E_{p}=N_{ij}^{\rm abs} \frac{L_{*} \delta t}{N_{p}} \ .
\end{equation}
The total energy emitted in a zone over the time interval $\delta t$ is: 
\begin{equation} 
\label{eq:E2}
E_{ij}^{\rm em} = 4\pi\delta t \int dV_{ij} \int \rho \kappa^{\rm a}
_{\lambda}B_{\lambda}(T_{ij}) d\lambda \; .
\end{equation}
Expressing Equation~(\ref{eq:E2}) in terms of  the Planck mean opacity $\kappa_{\rm P}(T)= {\pi \int \kappa^{\rm a}_{\lambda} B_{\lambda}(T) d\lambda }/(\sigma_{\rm B} T^{4})$, where $\sigma_{\rm B}$ is the Stefan-Boltzmann constant, and using the fact that $\int \rho \, dV_{ij}$ equals the mass $m_{ij}$ of grains in the zone $\{ij\}$, leads to
\begin{equation} 
\label{eq:E3}
 E_{ij}^{\rm em}= 4\, \delta t\, m_{ij} \kappa_{\rm P} (T_{ij}) \sigma_{\rm B} T_{ij}^{4} \; .
\end{equation}
To satisfy radiative equilibrium, the total absorbed energy in each zone must be reemitted within the same zone, and therefore we equate the expressions (\ref{eq:E1}) and~(\ref{eq:E3}) to obtain the following formula for the dust temperature within a zone:
\begin{equation}
\label{eq:E4}
T_{ij}^{4}=\frac{ N_{ij}^{\rm abs} L_{*}}{4N_{p} \kappa_{\rm P}(T_{ij})\sigma_{\rm B} m_{ij}} \ .
\end{equation} 
Equation~(\ref{eq:E4}) is implicit and therefore has to be solved iteratively for $T_{ij}$ each time a packet is absorbed. To save time, the Planck mean opacities can be precalculated for a range of temperatures, and then values needed in this equation can be interpolated from the precalculated files. After the packet has been absorbed, it is reemitted with a \emph{new} wavelength. This new wavelength is determined by the temperature correction scheme of \citet{BjorkmanWood01}, which we now summarize for the reader's convenience.

Imagine a packet is absorbed in the zone $\{ij\}$ and reemitted right away based on the 
local dust temperature $T^\prime_{ij}$.  When a subsequent packet is absorbed in the zone, the temperature increases to $T_{ij}=T^\prime_{ij} +\Delta T$.  It is clear that the first packet was reemitted with the wrong wavelength corresponding to $T_{ij} - \Delta T$; it should have been reemitted with more energy because the temperature is in fact higher. The implied additional monochromatic luminosity per unit mass, $\Delta{\cal{L}}_\lambda$, is given by
\begin{equation}
\label{eq:E5}  
\Delta {\cal{L}}_{\lambda}={\cal{L}}_{\lambda}-{\cal{L}}^{\prime}_{\lambda}
=\kappa^{\rm a}_{\lambda}[B_{\lambda}(T_{ij})- B_{\lambda}(T_{ij}-\Delta T)] \ ,
\end{equation}
which, when $\Delta T/T_{ij}$ is small (which can be ensured by sampling a sufficiently large number of packets), becomes  
\begin{equation} 
\label{eq:E6}
\Delta {\cal{L}}_{\lambda}=\kappa^{\rm a}_{\lambda} \Delta T \frac{dB_{\lambda}(T_{ij})}{dT} \ .
\end{equation}
To correct for the previously emitted spectrum, the reemitted wavelength is drawn from a probability 
distribution that is based on $\Delta{\cal{L}}_{\lambda}$ using the expression
\begin{equation} 
\label{eq:E7}
 \frac{k+0.5}{K}= \frac{\int_0 ^{\lambda^{re}_{k}} \sigma_{\lambda}^{a}\frac{dB_{\lambda}(T_{ij})}{dT}
  d\lambda}{\int_0^{\infty}\sigma^{a}_{\lambda} \frac{dB_{\lambda}(T_{ij})}{dT} d\lambda} \ , \quad\quad k=0,1,2.... K-1
\end{equation}
(cf. Equation(\ref{eq:B2})). This  correction is applied to every packet that is absorbed in a given zone. For computational efficiency, Equation~(\ref{eq:E7}) is evaluated for a range of different temperatures before the code is set to run. For each temperature there are $K$ probable reemitted wavelengths, and these values are tabulated into files and then read by and stored in the code.  This procedure yields values for the reemitted wavelength $\lambda^{\rm re}$ that are functions of the zone temperature and of a random integer $k=0,1...K-1$ (and thus are specified by 
$\lambda_{k}^{\rm re}(T_{ij})$).  Note that, even though the packet wavelength has changed, the reemitted packet still has an energy $E_{p}$, so energy is explicitly conserved.

\subsection{Spectral Energy Distribution}
\label{subsec:spectra}

Each packet that exits the system is binned into $N_{\lambda_{\rm bin}}$ wavelength bins and $N_{\rm inc}$ inclination bins, with indices $q$ and $l$, respectively:
\begin{equation}\label{eq:F1}
q=\frac{N_{\lambda_{\rm bin}} \lambda } {\lambda_{\rm max}} \ ,\quad l=N_{\rm inc} \cos{\theta}\; . 
\end{equation}
Here $\lambda_{\rm max}$ is chosen to be some maximum wavelength that the processed packets reach; for the region modeled in this analysis, $\lambda_{\rm max} \approx 1000\, \micron$. The specific flux per bin of the exiting packets is
\begin{equation} 
\label{eq:F2}
F_{\lambda}= \frac{N_{\rm ql} E_{p} N_{\rm inc}}{4\pi d^{2} \delta t \Delta \lambda}  \ ,  
\end{equation}
where $d$ is the distance to the observer and $N_{\rm ql}$ is the number of packets that occupy 
the $\{ql\}$ bin. The total flux from the star, $F_*=L_*/(4\pi d^2)$, can be rewritten with the help of Equation~(\ref{eq:B1}) as
\begin{equation}  
\label{eq:F3}
F_*=\frac{N_p E_{p}}{4\pi d^{2} \delta t } \ . 
\end{equation}
The effective width of each bin is $\Delta \lambda = {\lambda_{\rm max}}/{N_{\lambda_{\rm bin}}}$, which can be expressed, using Equation~(\ref{eq:F1}), as $\Delta \lambda = \lambda /q$. Taking the wavelength at the center of the bin and using Equation~(\ref{eq:F2}) and~(\ref{eq:F3}) yields a formula for the normalized SED flux:
\begin{equation}  
\label{eq:F4}
\frac{\lambda F_{\lambda}}{F_*}= (q+ 0.5) \frac{N_{\rm ql} N_{\rm inc}}{N_{p} }\  . 
\end{equation}

\bibliographystyle{apj}
\bibliography{mybib}

\begin{thebibliography}{86}
\expandafter\ifx\csname natexlab\endcsname\relax\def\natexlab#1{#1}\fi

\bibitem[{{Adams} \& {Shu}(1986)}]{AdamsShu86}
{Adams}, F.~C., \& {Shu}, F.~H. 1986, \apj, 308, 836

\bibitem[{{Akeson} {et~al.}(2005){Akeson}, {Walker}, {Wood}, {Eisner}, {Scire},
  {Penprase}, {Ciardi}, {van Belle}, {Whitney}, \& {Bjorkman}}]{AkesonEtc05}
{Akeson}, R.~L., {et~al.} 2005, \apj, 622, 440

\bibitem[{{Alecian} {et~al.}(2009){Alecian}, {Wade}, {Catala}, {Bagnulo},
  {B{\"o}hm}, {Bouret}, {Donati}, {Folsom}, {Grunhut}, \&
  {Landstreet}}]{AlecianEtc09}
{Alecian}, E., {et~al.} 2009, \mnras, 400, 354

\bibitem[{{Alexander} {et~al.}(2005){Alexander}, {Clarke}, \&
  {Pringle}}]{AlexanderEtc05}
{Alexander}, R.~D., {Clarke}, C.~J., \& {Pringle}, J.~E. 2005, \mnras, 358, 283

\bibitem[{{Barvainis}(1990)}]{Barvainis90}
{Barvainis}, R. 1990, \apj, 353, 419

\bibitem[{{Benisty} {et~al.}(2010){Benisty}, {Natta}, {Isella}, {Berger},
  {Massi}, {Le Bouquin}, {M{\'e}rand}, {Duvert}, {Kraus}, {Malbet}, {Olofsson},
  {Robbe-Dubois}, {Testi}, {Vannier}, \& {Weigelt}}]{BenistyEtc10}
{Benisty}, M., {et~al.} 2010, \aap, 511, A74

\bibitem[{{Berger}(2003)}]{Berger03}
{Berger}, J.-P. 2003, in EAS Publications Series, Vol.~6, EAS Publications
  Series, ed. {G.~Perrin \& F.~Malbet}, 23--+

\bibitem[{{Bjorkman} \& {Wood}(2001)}]{BjorkmanWood01}
{Bjorkman}, J.~E., \& {Wood}, K. 2001, \apj, 554, 615

\bibitem[{{Blandford} \& {Payne}(1982)}]{BP82}
{Blandford}, R.~D., \& {Payne}, D.~G. 1982, \mnras, 199, 883

\bibitem[{{Calvet} {et~al.}(2000){Calvet}, {Hartmann}, \&
  {Strom}}]{CalvetEtc00}
{Calvet}, N., {Hartmann}, L., \& {Strom}, S.~E. 2000, in Protostars and Planets
  IV, ed. V.~Mannings, A.~Boss, \& S.~Russell (Tucson: Univ. Arizona Press),
  377

\bibitem[{{Casse} \& {Ferreira}(2000)}]{CasseFerreira00}
{Casse}, F., \& {Ferreira}, J. 2000, \aap, 361, 1178

\bibitem[{{Chiang} \& {Goldreich}(1997)}]{ChiangGoldreich97}
{Chiang}, E.~I., \& {Goldreich}, P. 1997, \apj, 490, 368

\bibitem[{{Chiang} {et~al.}(2001){Chiang}, {Joung}, {Creech-Eakman}, {Qi},
  {Kessler}, {Blake}, \& {van Dishoeck}}]{ChiangEtc01}
{Chiang}, E.~I., {Joung}, M.~K., {Creech-Eakman}, M.~J., {Qi}, C., {Kessler},
  J.~E., {Blake}, G.~A., \& {van Dishoeck}, E.~F. 2001, \apj, 547, 1077

\bibitem[{{Chrysostomou} {et~al.}(2008){Chrysostomou}, {Bacciotti}, {Nisini},
  {Ray}, {Eisl{\"o}ffel}, {Davis}, \& {Takami}}]{ChrysostomouEtc08}
{Chrysostomou}, A., {Bacciotti}, F., {Nisini}, B., {Ray}, T.~P.,
  {Eisl{\"o}ffel}, J., {Davis}, C.~J., \& {Takami}, M. 2008, \aap, 482, 575

\bibitem[{{Chrysostomou} {et~al.}(2007){Chrysostomou}, {Lucas}, \&
  {Hough}}]{ChrysostomouEtc07}
{Chrysostomou}, A., {Lucas}, P.~W., \& {Hough}, J.~H. 2007, \nat, 450, 71

\bibitem[{{Coffey} {et~al.}(2008){Coffey}, {Bacciotti}, \&
  {Podio}}]{CoffeyEtc08}
{Coffey}, D., {Bacciotti}, F., \& {Podio}, L. 2008, \apj, 689, 1112

\bibitem[{{Corcoran} \& {Ray}(1998)}]{CorcoranRay98}
{Corcoran}, M., \& {Ray}, T.~P. 1998, \aap, 331, 147

\bibitem[{{Cranmer}(2009)}]{Cranmer09}
{Cranmer}, S.~R. 2009, \apj, 706, 824

\bibitem[{{Donehew} \& {Brittain}(2011)}]{DonehewBrittain11}
{Donehew}, B., \& {Brittain}, S. 2011, \aj, 141, 46

\bibitem[{{Dullemond} \& {Monnier}(2010)}]{DMReview}
{Dullemond}, C.~P., \& {Monnier}, J.~D. 2010, \araa, 48, 205 (DM10)

\bibitem[{{Edelson} \& {Malkan}(1986)}]{EdelsonMalkan86}
{Edelson}, R.~A., \& {Malkan}, M.~A. 1986, \apj, 308, 59

\bibitem[{{Eisner} {et~al.}(2010){Eisner}, {Monnier}, {Woillez}, {Akeson},
  {Millan-Gabet}, {Graham}, {Hillenbrand}, {Pott}, {Ragland}, \&
  {Wizinowich}}]{EisnerEtc10}
{Eisner}, J.~A., {et~al.} 2010, \apj, 718, 774

\bibitem[{{Elitzur} \& {Shlosman}(2006)}]{ElitzurShlosman06}
{Elitzur}, M., \& {Shlosman}, I. 2006, \apjl, 648, L101

\bibitem[{{Ferguson} {et~al.}(2005){Ferguson}, {Alexander}, {Allard}, {Barman},
  {Bodnarik}, {Hauschildt}, {Heffner-Wong}, \& {Tamanai}}]{FergusonEtc05}
{Ferguson}, J.~W., {Alexander}, D.~R., {Allard}, F., {Barman}, T., {Bodnarik},
  J.~G., {Hauschildt}, P.~H., {Heffner-Wong}, A., \& {Tamanai}, A. 2005, \apj,
  623, 585

\bibitem[{{Garcia Lopez} {et~al.}(2006){Garcia Lopez}, {Natta}, {Testi}, \&
  {Habart}}]{GarciaLopezEtc06}
{Garcia Lopez}, R., {Natta}, A., {Testi}, L., \& {Habart}, E. 2006, \aap, 459,
  837

\bibitem[{{Gorti} \& {Hollenbach}(2009)}]{GortiHollenbach09}
{Gorti}, U., \& {Hollenbach}, D. 2009, \apj, 690, 1539

\bibitem[{{Gueth} {et~al.}(2003){Gueth}, {Bachiller}, \&
  {Tafalla}}]{GuethEtc03}
{Gueth}, F., {Bachiller}, R., \& {Tafalla}, M. 2003, \aap, 401, L5

\bibitem[{{Hartigan} {et~al.}(1995){Hartigan}, {Edwards}, \&
  {Ghandour}}]{HartiganEtc95}
{Hartigan}, P., {Edwards}, S., \& {Ghandour}, L. 1995, \apj, 452, 736

\bibitem[{{Hillenbrand} {et~al.}(1992){Hillenbrand}, {Strom}, {Vrba}, \&
  {Keene}}]{HillenbrandEtc92}
{Hillenbrand}, L.~A., {Strom}, S.~E., {Vrba}, F.~J., \& {Keene}, J. 1992, \apj,
  397, 613

\bibitem[{{Hubeny}(1990)}]{Hubeny90}
{Hubeny}, I. 1990, \apj, 351, 632

\bibitem[{{Hubrig} {et~al.}(2011){Hubrig}, {Mikul{\'a}{\v s}ek},
  {Gonz{\'a}lez}, {Sch{\"o}ller}, {Ilyin}, {Cur{\'e}}, {Zejda}, {Cowley},
  {Elkin}, {Pogodin}, \& {Yudin}}]{HubrigEtc11}
{Hubrig}, S., {et~al.} 2011, \aap, 525, L4

\bibitem[{{Isella} \& {Natta}(2005)}]{IsellaNatta05}
{Isella}, A., \& {Natta}, A. 2005, \aap, 438, 899

\bibitem[{{Isella} {et~al.}(2006){Isella}, {Testi}, \& {Natta}}]{IsellaEtc06}
{Isella}, A., {Testi}, L., \& {Natta}, A. 2006, \aap, 451, 951

\bibitem[{{Ivezi{\'c}} \& {Elitzur}(1997)}]{IvezicElitzur97}
{Ivezi{\'c}}, {\v Z}., \& {Elitzur}, M. 1997, \mnras, 287, 799

\bibitem[{{Kartje} {et~al.}(1999){Kartje}, {K{\"o}nigl}, \&
  {Elitzur}}]{KartjeEtc99}
{Kartje}, J.~F., {K{\"o}nigl}, A., \& {Elitzur}, M. 1999, \apj, 513, 180

\bibitem[{{Kondratko} {et~al.}(2005){Kondratko}, {Greenhill}, \&
  {Moran}}]{KondratkoEtc05}
{Kondratko}, P.~T., {Greenhill}, L.~J., \& {Moran}, J.~M. 2005, \apj, 618, 618

\bibitem[{{K{\"o}nigl}(1996)}]{Konigl96}
{K{\"o}nigl}, A. 1996, in Lecture Notes in Physics, Berlin Springer Verlag,
  Vol. 465, Disks and Outflows Around Young Stars, ed. S.~{Beckwith},
  J.~{Staude}, A.~{Quetz}, \& A.~{Natta}, 282

\bibitem[{{K\"onigl} \& {Kartje}(1994)}]{KK94}
{K\"onigl}, A., \& {Kartje}, J.~F. 1994, \apj, 434, 446

\bibitem[{{K\"onigl} \& {Pudritz}(2000)}]{KoniglPudritz00}
{K\"onigl}, A., \& {Pudritz}, R.~E. 2000, in Protostars and Planets V, ed.
  V.~Mannings, A.~Boss, \& S.~Russell (Tucson: Univ. Arizona Press), 759

\bibitem[{{K{\"o}nigl} \& {Salmeron}(2011)}]{KoniglSalmeron11}
{K{\"o}nigl}, A., \& {Salmeron}, R. 2011, in Physical Processes in
  Circumstellar Disks around Young Stars, ed. P.~J.~V. {Garcia} (Chicago: Univ.
  Chicago Press), 283

\bibitem[{{Kraus} {et~al.}(2008){Kraus}, {Hofmann}, {Benisty}, {Berger},
  {Chesneau}, {Isella}, {Malbet}, {Meilland}, {Nardetto}, {Natta}, {Preibisch},
  {Schertl}, {Smith}, {Stee}, {Tatulli}, {Testi}, \& {Weigelt}}]{KrausEtc08}
{Kraus}, S., {et~al.} 2008, \aap, 489, 1157

\bibitem[{{Long} {et~al.}(2005){Long}, {Romanova}, \& {Lovelace}}]{LongEtc05}
{Long}, M., {Romanova}, M.~M., \& {Lovelace}, R.~V.~E. 2005, \apj, 634, 1214

\bibitem[{{Lucy}(1999)}]{Lucy99}
{Lucy}, L.~B. 1999, \aap, 344, 282

\bibitem[{{Meeus} {et~al.}(2001){Meeus}, {Waters}, {Bouwman}, {van den Ancker},
  {Waelkens}, \& {Malfait}}]{MeeusEtc01}
{Meeus}, G., {Waters}, L.~B.~F.~M., {Bouwman}, J., {van den Ancker}, M.~E.,
  {Waelkens}, C., \& {Malfait}, K. 2001, \aap, 365, 476

\bibitem[{{Mendigut{\'{\i}}a} {et~al.}(2011){Mendigut{\'{\i}}a}, {Calvet},
  {Montesinos}, {Mora}, {Muzerolle}, {Eiroa}, {Oudmaijer}, \&
  {Mer{\'{\i}}n}}]{MendigutiaEtc11}
{Mendigut{\'{\i}}a}, I., {Calvet}, N., {Montesinos}, B., {Mora}, A.,
  {Muzerolle}, J., {Eiroa}, C., {Oudmaijer}, R.~D., \& {Mer{\'{\i}}n}, B. 2011,
  \aap, 535, A99

\bibitem[{{Millan-Gabet} {et~al.}(2007){Millan-Gabet}, {Malbet}, {Akeson},
  {Leinert}, {Monnier}, \& {Waters}}]{MillanGabetEtc07}
{Millan-Gabet}, R., {Malbet}, F., {Akeson}, R., {Leinert}, C., {Monnier}, J.,
  \& {Waters}, R. 2007, in Protostars and Planets V, ed. B.~Reipurth,
  D.~Jewitt, \& K.~Keil (Tucson: Univ. Arizona Press), 539

\bibitem[{{Monnier} {et~al.}(2006){Monnier}, {Berger}, {Millan-Gabet}, {Traub},
  {Schloerb}, {Pedretti}, {Benisty}, {Carleton}, {Haguenauer}, {Kern},
  {Labeye}, {Lacasse}, {Malbet}, {Perraut}, {Pearlman}, \&
  {Zhao}}]{MonnierEtc06}
{Monnier}, J.~D., {et~al.} 2006, \apj, 647, 444

\bibitem[{{Monnier} \& {Millan-Gabet}(2002)}]{MonnierMillanGabet02}
{Monnier}, J.~D., \& {Millan-Gabet}, R. 2002, \apj, 579, 694

\bibitem[{{Monnier} {et~al.}(2005){Monnier}, {Millan-Gabet}, {Billmeier},
  {Akeson}, {Wallace}, {Berger}, {Calvet}, {D'Alessio}, {Danchi}, {Hartmann},
  {Hillenbrand}, {Kuchner}, {Rajagopal}, {Traub}, {Tuthill}, {Boden}, {Booth},
  {Colavita}, {Gathright}, {Hrynevych}, {Le Mignant}, {Ligon}, {Neyman},
  {Swain}, {Thompson}, {Vasisht}, {Wizinowich}, {Beichman}, {Beletic},
  {Creech-Eakman}, {Koresko}, {Sargent}, {Shao}, \& {van Belle}}]{MonnierEtc05}
{Monnier}, J.~D., {et~al.} 2005, \apj, 624, 832

\bibitem[{{Muzerolle} {et~al.}(2003){Muzerolle}, {Calvet}, {Hartmann}, \&
  {D'Alessio}}]{MuzerolleEtc03}
{Muzerolle}, J., {Calvet}, N., {Hartmann}, L., \& {D'Alessio}, P. 2003, \apjl,
  597, L149

\bibitem[{{Muzerolle} {et~al.}(2004){Muzerolle}, {D'Alessio}, {Calvet}, \&
  {Hartmann}}]{MuzerolleEtc04}
{Muzerolle}, J., {D'Alessio}, P., {Calvet}, N., \& {Hartmann}, L. 2004, \apj,
  617, 406

\bibitem[{{Muzerolle} {et~al.}(2009){Muzerolle}, {Flaherty}, {Balog}, {Furlan},
  {Smith}, {Allen}, {Calvet}, {D'Alessio}, {Megeath}, {Muench}, {Rieke}, \&
  {Sherry}}]{MuzerolleEtc09}
{Muzerolle}, J., {et~al.} 2009, \apjl, 704, L15

\bibitem[{{Natta} {et~al.}(2001){Natta}, {Prusti}, {Neri}, {Wooden}, {Grinin},
  \& {Mannings}}]{NattaEtc01}
{Natta}, A., {Prusti}, T., {Neri}, R., {Wooden}, D., {Grinin}, V.~P., \&
  {Mannings}, V. 2001, \aap, 371, 186

\bibitem[{{Neckel} \& {Staude}(1995)}]{NeckelStaude95}
{Neckel}, T., \& {Staude}, H.~J. 1995, \apj, 448, 832

\bibitem[{{Nenkova} {et~al.}(2008){Nenkova}, {Sirocky}, {Nikutta},
  {Ivezi{\'c}}, \& {Elitzur}}]{NenkovaEtc08}
{Nenkova}, M., {Sirocky}, M.~M., {Nikutta}, R., {Ivezi{\'c}}, {\v Z}., \&
  {Elitzur}, M. 2008, \apj, 685, 160

\bibitem[{{Nisini} {et~al.}(2005){Nisini}, {Bacciotti}, {Giannini}, {Massi},
  {Eisl{\"o}ffel}, {Podio}, \& {Ray}}]{NisiniEtc05}
{Nisini}, B., {Bacciotti}, F., {Giannini}, T., {Massi}, F., {Eisl{\"o}ffel},
  J., {Podio}, L., \& {Ray}, T.~P. 2005, \aap, 441, 159

\bibitem[{{Nisini} {et~al.}(1995){Nisini}, {Milillo}, {Saraceno}, \&
  {Vitali}}]{NisiniEtc95}
{Nisini}, B., {Milillo}, A., {Saraceno}, P., \& {Vitali}, F. 1995, \aap, 302,
  169

\bibitem[{{O'Brien} {et~al.}(2003){O'Brien}, {Garcia}, {Ferreira}, {Cabrit}, \&
  {Binette}}]{ObrienEtc03}
{O'Brien}, D., {Garcia}, P., {Ferreira}, J., {Cabrit}, S., \& {Binette}, L.
  2003, \apss, 287, 129

\bibitem[{{Pesenti} {et~al.}(2004){Pesenti}, {Dougados}, {Cabrit}, {Ferreira},
  {Casse}, {Garcia}, \& {O'Brien}}]{PesentiEtc04}
{Pesenti}, N., {Dougados}, C., {Cabrit}, S., {Ferreira}, J., {Casse}, F.,
  {Garcia}, P., \& {O'Brien}, D. 2004, \aap, 416, L9

\bibitem[{{Pollack} {et~al.}(1994){Pollack}, {Hollenbach}, {Beckwith},
  {Simonelli}, {Roush}, \& {Fong}}]{PollackEtc94}
{Pollack}, J.~B., {Hollenbach}, D., {Beckwith}, S., {Simonelli}, D.~P.,
  {Roush}, T., \& {Fong}, W. 1994, \apj, 421, 615

\bibitem[{{Pudritz} \& {Norman}(1983)}]{PudritzNorman83}
{Pudritz}, R.~E., \& {Norman}, C.~A. 1983, \apj, 274, 677

\bibitem[{{Pudritz} {et~al.}(2007){Pudritz}, {Ouyed}, {Fendt}, \&
  {Brandenburg}}]{PudritzEtc07}
{Pudritz}, R.~E., {Ouyed}, R., {Fendt}, C., \& {Brandenburg}, A. 2007, in
  Protostars and Planets V, ed. B.~Reipurth, D.~Jewitt, \& K.~Keil (Tucson:
  Univ. Arizona Press), 277

\bibitem[{{Ray} {et~al.}(2007){Ray}, {Dougados}, {Bacciotti}, {Eisl{\"o}ffel},
  \& {Chrysostomou}}]{RayEtc07}
{Ray}, T., {Dougados}, C., {Bacciotti}, F., {Eisl{\"o}ffel}, J., \&
  {Chrysostomou}, A. 2007, in Protostars and Planets V, ed. B.~Reipurth,
  D.~Jewitt, \& K.~Keil (Tucson: Univ. Arizona Press), 231

\bibitem[{{Romanova} {et~al.}(2009){Romanova}, {Ustyugova}, {Koldoba}, \&
  {Lovelace}}]{RomanovaEtc09}
{Romanova}, M.~M., {Ustyugova}, G.~V., {Koldoba}, A.~V., \& {Lovelace},
  R.~V.~E. 2009, \mnras, 399, 1802

\bibitem[{{Safier}(1993{\natexlab{a}})}]{Safier93a}
{Safier}, P.~N. 1993{\natexlab{a}}, \apj, 408, 115

\bibitem[{{Safier}(1993{\natexlab{b}})}]{Safier93b}
---. 1993{\natexlab{b}}, \apj, 408, 148

\bibitem[{{Salmeron} \& {Ireland}(2012)}]{SalmeronIreland12}
{Salmeron}, R., \& {Ireland}, T.~R. 2012, Earth and Planetary Science Letters,
  327, 61

\bibitem[{{Salmeron} {et~al.}(2007){Salmeron}, {K{\"o}nigl}, \&
  {Wardle}}]{SalmeronEtc07}
{Salmeron}, R., {K{\"o}nigl}, A., \& {Wardle}, M. 2007, \mnras, 375, 177

\bibitem[{{Shang} {et~al.}(2002){Shang}, {Glassgold}, {Shu}, \&
  {Lizano}}]{ShangEtc02}
{Shang}, H., {Glassgold}, A.~E., {Shu}, F.~H., \& {Lizano}, S. 2002, \apj, 564,
  853

\bibitem[{{Shu} {et~al.}(2000){Shu}, {Najita}, {Shang}, \& {Li}}]{ShuEtc00}
{Shu}, F.~H., {Najita}, J.~R., {Shang}, H., \& {Li}, Z.-Y. 2000, in Protostars
  and Planets IV, ed. V.~Mannings, A.~Boss, \& S.~Russell (Tucson: Univ.
  Arizona Press), 789

\bibitem[{{Sitko} {et~al.}(2008){Sitko}, {Carpenter}, {Kimes}, {Wilde},
  {Lynch}, {Russell}, {Rudy}, {Mazuk}, {Venturini}, {Puetter}, {Grady},
  {Polomski}, {Wisnewski}, {Brafford}, {Hammel}, \& {Perry}}]{SitkoEtc08}
{Sitko}, M.~L., {et~al.} 2008, \apj, 678, 1070

\bibitem[{{Smith} {et~al.}(2005){Smith}, {Bally}, {Shuping}, {Morris}, \&
  {Kassis}}]{SmithEtc05}
{Smith}, N., {Bally}, J., {Shuping}, R.~Y., {Morris}, M., \& {Kassis}, M. 2005,
  \aj, 130, 1763

\bibitem[{{Tambovtseva} \& {Grinin}(2008)}]{TambovtsevaGrinin08}
{Tambovtseva}, L.~V., \& {Grinin}, V.~P. 2008, Astronomy Letters, 34, 231

\bibitem[{{Tannirkulam} {et~al.}(2007){Tannirkulam}, {Harries}, \&
  {Monnier}}]{TannirkulamEtc07}
{Tannirkulam}, A., {Harries}, T.~J., \& {Monnier}, J.~D. 2007, \apj, 661, 374

\bibitem[{{Tannirkulam} {et~al.}(2008b){Tannirkulam}, {Monnier}, {Harries},
  {Millan-Gabet}, {Zhu}, {Pedretti}, {Ireland}, {Tuthill}, {ten Brummelaar},
  {McAlister}, {Farrington}, {Goldfinger}, {Sturmann}, {Sturmann}, \&
  {Turner}}]{TannirkulamEtc08b}
{Tannirkulam}, A., {et~al.} 2008b, \apj, 689, 513

\bibitem[{{Tannirkulam} {et~al.}(2008a){Tannirkulam}, {Monnier},
  {Millan-Gabet}, {Harries}, {Pedretti}, {ten Brummelaar}, {McAlister},
  {Turner}, {Sturmann}, \& {Sturmann}}]{TannirkulamEtc08a}
---. 2008a, \apjl, 677, L51

\bibitem[{{Tatulli} {et~al.}(2007){Tatulli}, {Isella}, {Natta}, {Testi},
  {Marconi}, {Malbet}, {Stee}, {Petrov}, {Millour}, {Chelli}, {Duvert},
  {Antonelli}, {Beckmann}, {Bresson}, {Dugu{\'e}}, {Gennari}, {Gl{\"u}ck},
  {Kern}, {Lagarde}, {Le Coarer}, {Lisi}, {Perraut}, {Puget}, {Rantakyr{\"o}},
  {Robbe-Dubois}, {Roussel}, {Weigelt}, {Zins}, {Accardo}, {Acke}, {Agabi},
  {Altariba}, {Arezki}, {Aristidi}, {Baffa}, {Behrend}, {Bl{\"o}cker},
  {Bonhomme}, {Busoni}, {Cassaing}, {Clausse}, {Colin}, {Connot},
  {Delboulb{\'e}}, {Domiciano de Souza}, {Driebe}, {Feautrier}, {Ferruzzi},
  {Forveille}, {Fossat}, {Foy}, {Fraix-Burnet}, {Gallardo}, {Giani}, {Gil},
  {Glentzlin}, {Heiden}, {Heininger}, {Hernandez Utrera}, {Hofmann}, {Kamm},
  {Kiekebusch}, {Kraus}, {Le Contel}, {Le Contel}, {Lesourd}, {Lopez}, {Lopez},
  {Magnard}, {Mars}, {Martinot-Lagarde}, {Mathias}, {M{\`e}ge}, {Monin},
  {Mouillet}, {Mourard}, {Nussbaum}, {Ohnaka}, {Pacheco}, {Perrier}, {Rabbia},
  {Rebattu}, {Reynaud}, {Richichi}, {Robini}, {Sacchettini}, {Schertl},
  {Sch{\"o}ller}, {Solscheid}, {Spang}, {Stefanini}, {Tallon}, {Tallon-Bosc},
  {Tasso}, {Vakili}, {von der L{\"u}he}, {Valtier}, {Vannier}, \&
  {Ventura}}]{TatulliEtc07}
{Tatulli}, E., {et~al.} 2007, \aap, 464, 55

\bibitem[{{Tuthill} {et~al.}(2001){Tuthill}, {Monnier}, \&
  {Danchi}}]{TuthillEtc01}
{Tuthill}, P.~G., {Monnier}, J.~D., \& {Danchi}, W.~C. 2001, \nat, 409, 1012

\bibitem[{{Vink} {et~al.}(2005){Vink}, {Drew}, {Harries}, {Oudmaijer}, \&
  {Unruh}}]{VinkEtc05}
{Vink}, J.~S., {Drew}, J.~E., {Harries}, T.~J., {Oudmaijer}, R.~D., \& {Unruh},
  Y. 2005, \mnras, 359, 1049

\bibitem[{{Vinkovi{\'c}} {et~al.}(2006){Vinkovi{\'c}}, {Ivezi{\'c}},
  {Jurki{\'c}}, \& {Elitzur}}]{VinkovicEtc06}
{Vinkovi{\'c}}, D., {Ivezi{\'c}}, {\v Z}., {Jurki{\'c}}, T., \& {Elitzur}, M.
  2006, \apj, 636, 348

\bibitem[{{Vinkovi{\'c}} \& {Jurki{\'c}}(2007)}]{VinkovicJurkic07}
{Vinkovi{\'c}}, D., \& {Jurki{\'c}}, T. 2007, \apj, 658, 462

\bibitem[{{Wade} {et~al.}(2011){Wade}, {Alecian}, {Grunhut}, {Catala},
  {Bagnulo}, {Folsom}, \& {Landstreet}}]{WadeEtc2011}
{Wade}, G.~A., {Alecian}, E., {Grunhut}, J., {Catala}, C., {Bagnulo}, S.,
  {Folsom}, C.~P., \& {Landstreet}, J.~D. 2011, in Astronomical Society of the
  Pacific Conference Series, Vol. 449, Astronomical Society of the Pacific
  Conference Series, ed. P.~{Bastien}, N.~{Manset}, D.~P. {Clemens}, \&
  N.~{St-Louis}, 262

\bibitem[{{Wassell} {et~al.}(2006){Wassell}, {Grady}, {Woodgate}, {Kimble}, \&
  {Bruhweiler}}]{WassellEtc06}
{Wassell}, E.~J., {Grady}, C.~A., {Woodgate}, B., {Kimble}, R.~A., \&
  {Bruhweiler}, F.~C. 2006, \apj, 650, 985

\bibitem[{{Weigelt} {et~al.}(2011){Weigelt}, {Grinin}, {Groh}, {Hofmann},
  {Kraus}, {Miroshnichenko}, {Schertl}, {Tambovtseva}, {Benisty}, {Driebe},
  {Lagarde}, {Malbet}, {Meilland}, {Petrov}, \& {Tatulli}}]{WeigeltEtc11}
{Weigelt}, G., {et~al.} 2011, \aap, 527, A103

\bibitem[{{Weingartner} \& {Draine}(2001)}]{WeingartnerDraine01}
{Weingartner}, J.~C., \& {Draine}, B.~T. 2001, \apj, 548, 296

\bibitem[{{Wisniewski} {et~al.}(2008){Wisniewski}, {Clampin}, {Grady},
  {Ardila}, {Ford}, {Golimowski}, {Illingworth}, \& {Krist}}]{WisniewskiEtc08}
{Wisniewski}, J.~P., {Clampin}, M., {Grady}, C.~A., {Ardila}, D.~R., {Ford},
  H.~C., {Golimowski}, D.~A., {Illingworth}, G.~D., \& {Krist}, J.~E. 2008,
  \apj, 682, 548

\end{thebibliography}

\end{document}